\documentclass[aps,10pt,pra,onecolumn,groupedaddress, floatfix, superscriptaddress,showpacs, showkeys,amsfonts]{revtex4-2}
\usepackage{amsfonts}
\usepackage{amsmath}
\usepackage{amssymb, setspace}
\usepackage{graphicx}
\graphicspath{{./}}
\usepackage{subfigure}
\usepackage{wrapfig}
\usepackage[%
  colorlinks=true,
  urlcolor=blue,
  linkcolor=blue,
  citecolor=blue
]{hyperref}
\usepackage[all]{hypcap}
\usepackage{color}
\usepackage[utf8]{inputenc}
\usepackage{dsfont}

\usepackage{comment}
\usepackage{soul}
\usepackage{orcidlink}
\usepackage{natbib}
\usepackage{braket}

\newcommand{\riverlane}{Riverlane, Cambridge, UK}

\begin{document}
\title{Discovering correlated error mechanisms: data-driven augmentation of detector error models improving decoding}

\author{Zohar Schwartzman-Nowik\orcidlink{0000-0003-3804-4933}}
\email{zohar.schwartzmannowik@riverlane.com}
\affiliation{\riverlane}

\author{Matthew P. Stafford\orcidlink{0000-0001-9938-3854}}
\affiliation{\riverlane}

\author{Boris M. Varbanov\orcidlink{0000-0001-7124-8933}}
\affiliation{\riverlane}

\author{Joan Camps\orcidlink{0000-0003-3896-7922}}
\affiliation{\riverlane}

\begin{abstract}
Accurate decoder priors are essential for high-performance quantum error correction. Theoretical noise models provide a useful starting point but inevitably omit physically relevant error mechanisms, most notably correlated error mechanisms, which appear consistently in experimental error budgets yet are absent from standard models. We introduce a data-driven method for augmenting detector error models (DEMs) by discovering entirely new error mechanisms directly from experimental detector-firing statistics, going beyond most existing approaches that merely refine the probabilities of already-known mechanisms. Our method generates candidate mechanisms by combining pairs of existing mechanisms as building blocks, then filters and validates them using pairwise and higher-order detector correlations. Crucially, this construction provides each newly discovered mechanism with a well-defined logical effect derived from the constituent error mechanisms, enabling its direct use in decoding. Furthermore, to preserve compatibility with the decoder supported by the original code and noise model, each newly discovered mechanism is accompanied by a natural decomposition into lower-order components. This is particularly valuable in settings that rely on correlated graph-based decoders such as correlated matching. We evaluate the method on surface-code simulations with correlated noise and correlated-matching decoding, demonstrating that, in a realistic near-term noise scenario, DEM augmentation increases the $\Lambda$ parameter by $20\%$ across code distances 7-15 and reduces the logical error rate by $75\%$ at code distance 15 relative to the unaugmented DEM. We further validate the method on experimental data from Google's superconducting device, where the augmented DEM increases the $\Lambda$ parameter by $9\%$ across code distances 5-7 and yields a reduction of $14\%$ in logical error rate over the baseline for distance 7. To our knowledge, this is the first data-driven DEM augmentation method that produces decomposed hyperedges with assigned logical effects, and we demonstrate its advantage in both controlled simulations and on real superconducting hardware.
\end{abstract}

\maketitle

\section{Introduction}

Quantum computers hold the promise of solving computational problems that are intractable for classical computers. Realizing this potential, however, requires overcoming the challenge posed by noise in quantum hardware. Quantum error correction (QEC) addresses this challenge by encoding quantum information redundantly across many physical qubits, allowing errors to be detected and corrected without disturbing the encoded information. Recent experiments have operated below-threshold in both the surface code \cite{google2024quantum} and the color code \cite{Lacroix2025scaling, Lavasani2026quantum}, with logical error rates suppressed as code distance increases, providing evidence that fault-tolerant quantum computation \cite{shor1996fault,gottesman1998theory} is within reach. The performance of a given QEC protocol depends on both the physical device characteristics and the decoder, a classical algorithm that uses measured syndrome data to infer an appropriate logical correction. Decoding accuracy is determined by the choice of decoding algorithm, as different decoders applied to the same experimental data can yield significantly different logical error rates \cite{Lavasani2026quantum, google2024quantum}, and, for a given decoding algorithm, how faithfully its model of the device noise represents the errors that actually occur. 

The decoder's noise model is typically encoded in a detector error model (DEM) \cite{gidney2021stim, derks2025designing}, which contains a set of independent error mechanisms, each specified by the detectors and logical observables it flips and its probability of occurrence. We refer to these error mechanisms and their probabilities as the \emph{decoder priors}. In practice, decoder priors are often derived from theoretical noise models \cite{gidney2021fault,gidney2022benchmarking,gidney2024inplace} that are typically straightforward to construct but are idealized descriptions of hardware. Real devices exhibit inhomogeneous noise, where error rates and biases vary across qubits, and may contain error mechanisms that are absent from standard models. A prominent example, frequently observed experimentally, is crosstalk \cite{google2023suppressing,google2024quantum,mundada2019suppression,sarovar2020detecting}, which introduces potentially distance-reducing correlations between otherwise distinct error processes. Accurately characterizing these error sources and including them within the DEM is vital to improving the performance of QEC protocols.

Fortunately, the measurement data collected during QEC provides a rich source of information about the underlying noise processes. A substantial body of work has used this data to improve decoder priors of an existing DEM \cite{spitz2018adaptive,google2021exponential,remm2026experimentally,blume2025estimating,takou2025estimating,wang2023dgr,sivak2024optimization}. These methods aim to refine the probabilities associated with error mechanisms already present in a base theoretical model. As they are restricted to the structure of the base model, they do not account for entirely new error mechanisms, such as crosstalk, that are absent from it. A small number of recent works have attempted to identify new error mechanisms directly from experimental data \cite{arms2025estimating,ye2026reconstruction}. However, these approaches do not assign the logical effects that the decoder requires to determine a correction. Furthermore, these methods introduce new error mechanisms without a valid decomposition into existing mechanisms present in the DEM. In the case of the surface code, introducing higher-order error mechanisms that flip more than two detectors without a valid decomposition means efficient graph-based decoders such as minimum-weight perfect matching (MWPM) \cite{dennis2002topological, fowler2012surface, edmonds1965paths} and union-find (UF) \cite{delfosse2021almost} may no longer be used. 

In this work, we develop a method that discovers new correlated error mechanisms directly from experimental data and assigns each one both its logical effects and a valid decomposition into existing error mechanisms. Our method takes a DEM obtained from a theoretical model and uses preexisting error mechanisms as ``building blocks'' to generate new candidate mechanisms. This avoids an exhaustive search over all possible detector subsets, a combinatorially intractable task. Candidates are first screened using fast pairwise detector correlations, and those that pass are subsequently validated through an explicit higher-order correlation calculation. By construction, the augmented DEM remains compatible with any decoder that accepts the original DEM, including graph-based decoders for the surface code, while decoders that exploit the decomposition structure, such as correlated matching, can additionally make use of the correlations the new mechanisms introduce.

We evaluate our proposed augmentation method by comparing the performance of correlated PyMatching \cite{higgott2022pymatching,higgott2025sparse,fowler2013optimal} with and without DEM augmentation on both simulated data and experimental data from Google's below-threshold surface-code quantum memory experiment \cite{google2024quantum, google2024zenodo}. For simulated data, we add a correlated two-qubit $XX$ error mechanism on nearby pairs of data qubits with probability $p_{\mathrm{corr}} = 3\times10^{-4}$ to an SI1000 noise model \cite{gidney2021fault,gidney2022benchmarking,gidney2024inplace} with noise parameter $p = 10^{-3}$. We build a baseline, unaugmented DEM for decoding using the SI1000 model alone. Using the augmented DEM increases $\Lambda$ (see Subsection \ref{subsec:logical error rate and lambda}), fitted over code distances 7-15, by $20\%$ and reduces the logical error rate at distance 15 by $75\%$ relative to using the baseline DEM. On experimental data, our baseline is the unaugmented SI1000 DEM with weights optimized via reinforcement learning (RL), as released by Google alongside their experimental data \cite{google2024quantum, google2024zenodo}. The augmented DEM reduces the logical error rate by $14\%$ over the baseline vanilla DEM at distance 7, and increases $\Lambda$ by $9\%$ over the baseline vanilla DEM for distances 5-7.

The remainder of this paper is organized as follows. Section \ref{sec:preliminaries} reviews the background needed to understand the methods and results presented throughout the paper. Section \ref{sec:augmentation method} then describes the proposed augmentation method, which forms the core contribution of this work. In Section \ref{sec:results}, we evaluate the method on the surface code and demonstrate the improvements it provides to decoding accuracy. Finally, Section \ref{sec:summary} concludes the paper with a summary of the main findings and a discussion of promising directions for future research.

\section{Preliminaries}\label{sec:preliminaries}

\subsection{Syndrome extraction, decoding, and detector error models}\label{subsec:syndrome_extraction}

A quantum error-correction code encodes logical information into a subspace of a larger physical Hilbert space, protecting it against a set of possible physical errors. Codes are defined by their set of \emph{stabilizer operators}, and the joint $+1$ eigenspace of the stabilizers is known as the codespace. Individual errors may take the encoded state out of the codespace into an error space. In addition, the error may also flip the value of an encoded logical observable, a fixed product of measurement outcomes corresponding to a representative of the codes's logical operator. To extract information about which errors may have occurred, the stabilizers of the code are indirectly measured using a set of \emph{ancilla qubits}. By measuring the stabilizers of the code, we reveal information about which error space the system is in without disturbing the encoded logical state. This measurement process, known as syndrome extraction, may itself be noisy and therefore must be repeated over multiple \emph{rounds}. The complete set of measurement data obtained by performing multiple rounds of syndrome extraction is called the \emph{syndrome}.

Rather than operating directly on raw syndrome measurements, modern decoders typically use \emph{detectors}: predefined combinations of measurement outcomes whose values are deterministic in the absence of noise. Deviations from the expected detector values, known as \emph{detection events}, provide a sparse and informative representation of the underlying error processes. We refer to the detectors on which such an event occurs as \emph{triggered}. Given an observed pattern of detection events, the role of the decoder is to infer a correction (if any) to be applied to the logical observables.

To infer a correction, the decoder requires a model of the possible errors and their corresponding likelihoods. Typically, this information is contained within a detector error model (DEM) \cite{gidney2021stim, derks2025designing}. Each entry in the DEM corresponds to an independent error mechanism, containing the set of detectors triggered by the error, the logical observables flipped by the error, and the error's probability of occurrence. We refer to this set of error mechanisms as the decoder priors, as they contain our preexisting knowledge of noise present in the hardware. 

The DEM may be equivalently represented as a weighted decoding hypergraph. In this representation, vertices correspond to detectors and hyperedges map to error mechanisms. Each hyperedge is connected to the set of detectors that the error mechanism triggers. The error mechanisms may be categorized by the number of detectors they trigger. If an error mechanism triggers only a single detector, we label it as a \emph{singleton}. If the mechanism triggers exactly two detectors, it forms an \emph{edge} in the decoding hypergraph. Finally, if the mechanism triggers three or more detectors, we say the mechanism forms a \emph{higher-order hyperedge} where the order corresponds to the number of detectors triggered by the error.

\subsection{Hyperedge decomposition}\label{subsec:hyperedge_decomposition}

If the DEM contains only singletons and edges, the decoding problem can be represented with a graph. Vertices of the graph correspond to detectors and edges map to error mechanisms. Singletons are made into edges by adding in a boundary vertex to the decoding graph. For graph-based decoding problems, efficient decoders such as MWPM and UF may be used \cite{fowler2013optimal,higgott2025sparse, dennis2002topological}. Therefore, recovering a graph-like decoding problem is a natural goal whenever possible. We note that the error-correcting code, circuit, and noise model all contribute to whether a DEM contains higher-order hyperedges. 

The \emph{surface code} \cite{kitaev2003fault,fowler2012surface} is a representative example. Qubits are arranged on a 2D square lattice where stabilizers are measured locally between neighboring qubits, see Figure \ref{fig:crosstalk types}. Here, a single-qubit $X$ or $Z$ error will trigger either a single detector at the boundary of the code or a pair of detectors in the bulk, resulting in a graph-based decoding problem. However, some error mechanisms will form higher-order hyperedges. For instance, a single-qubit $Y$ error in the bulk will trigger four detectors (a pair of $X$ and a pair of $Z$ type detectors), forming an order-4 hyperedge. Cross-basis two-qubit gate errors (for example, an $XZ$ error) can also produce hyperedges. Finally, correlated noise can generate an even wider variety of hyperedges, such as those where all detector events are of the same type (e.g., a single correlated Pauli $XX$ error may trigger four $Z$-type detection events).

To decode with a DEM containing hyperedges with a graph-based decoder, the hyperedges must first be \emph{decomposed} into edges and singletons. For some codes with inherently higher-order stabilizers, this decomposition is not possible. For codes where the basic syndrome measurements are lower order, such as the surface code, decomposition into a graph may be possible but depends on the structure of the circuit and noise model. To decompose a hyperedge, one needs to find a set of existing edges and singletons contained within the DEM that combine to match the detectors and observables contained within the original hyperedge. In this work, we use the decomposition functionality within Stim \cite{gidney2021stim}. For example, a hyperedge corresponding to a single-qubit $Y$ error may be decomposed into $X$ and $Z$ errors on the same qubit. For a correlated $XX$ error acting on separate qubits, the hyperedge may be decomposed into the product of two single-qubit $X$ errors. In this work, we call a hyperedge with a valid decomposition a \emph{decomposed hyperedge}. Otherwise, it is an \emph{undecomposed hyperedge}.

\subsection{Correlated decoding}\label{subsec:correlated_decoding}

While decomposing hyperedges into edges and singletons ensures the DEM is compatible with a graph-based decoder, it comes at a cost to accuracy. A standard single-pass decoder treats each component of the decomposed hyperedge independently, discarding the correlation between detectors that the original undecomposed hyperedge encoded. To recover this lost correlation while still benefiting from the efficiency of graph-based decoding, we use a correlated (or two-pass) matching approach \cite{fowler2013optimal,higgott2025sparse}, which decodes in two (or more) passes. The first pass runs a standard matching decoder to produce an initial error candidate. Each edge in this candidate that belongs to a decomposed hyperedge is used to update the probabilities of the other components of that hyperedge via a conditional reweighting map described in Appendix \ref{app:adaptivity map}. This allows information from one edge to influence other edges contained within the same hyperedge. A second decoding pass is then performed using these updated probabilities to find a logical correction \cite{fowler2013optimal,higgott2025sparse}.

For the surface code (and more generally CSS codes \cite{calderbank1996good, steane1996error}), each stabilizer is a product of either $X$ or $Z$ operators, so the detectors naturally split into $X$-type and $Z$-type. In the absence of cross-basis noise, each error mechanism only triggers detectors of a single type, and the decoding problem separates into two independent subgraphs. Cross-basis errors, such as single-qubit $Y$ errors or two-qubit $XZ$ errors, couple the two subgraphs into a single decoding hypergraph. These errors naturally decompose into mechanisms that contain one component in each subgraph. A pass on one subgraph may then be used to conditionally update the probabilities on the other. In a $Z$-basis quantum memory experiment, for example, performing a first pass over the $X$-type subgraph identifies likely $Y$ errors, and their components in the $Z$-type subgraph, which determines the logical outcome for the experiment, are then reweighted. For same-type correlated errors, such as the joint $XX$ errors considered in this work, both components of the decomposed hyperedge live in the same subgraph and no information is gained by performing a pass over the partner subgraph. In this case, the first decoding pass and reweighting must be performed on the same subgraph. Since the noise we consider contains both cross-basis and same-type correlations, we perform the first pass of correlated PyMatching over both subgraphs throughout this work.

\subsection{Logical error rate and $\Lambda$}
\label{subsec:logical error rate and lambda}

To evaluate decoding performance, we use the standard metrics of logical error probability, logical error rate, and $\Lambda$. Given a code distance $d$ and a fixed number of syndrome extraction rounds $r$, we repeatedly sample noisy syndromes, decode each one, and compare the decoder's predicted logical outcome to the true logical outcome. The \emph{logical error probability}, denoted $p_{L}(d,r)$, is the fraction of such trials in which the decoder's prediction disagrees with the true logical outcome.

Since the number of opportunities for a logical error to accumulate grows with the number of rounds, it is often more informative to report a per-round quantity. Assuming logical errors occur independently at a constant rate per round, with negligible state preparation and measurement (SPAM) contributions, the \emph{logical error rate (LER)} $\epsilon(d)$ \cite{google2024quantum} is the logical error probability per round and is obtained from $p_{L}(d,r)$ via 
\begin{equation}
    p_{L}(d,r)=\frac{1}{2}\left(1-\left(1-2\epsilon(d)\right)^{r}\right),\label{eq:ler by ler per round}
\end{equation}
which can be inverted to express $\epsilon(d)$ in terms of the
observed logical error probability,
\begin{equation}
\epsilon(d)=\frac{1}{2}\left(1-\left(1-2p_{L}(d,r)\right)^{1/r}\right),
\end{equation}
which allows for meaningful comparisons between experiments run with different numbers of rounds. In the regime where $p_{L}(d,r)\ll1$, the exact expression above can be simplified. Expanding to first order in $p_{L}(d,r)$ gives
\begin{equation}
\epsilon(d)\approx\frac{p_{L}(d,r)}{r}.\label{eq:ler per round by ler}
\end{equation}
Whenever logical errors are rare events, which is typically the case in the below-threshold regime for large enough distance and a modest number of rounds, the logical error rate can be well-approximated by Eq.~\eqref{eq:ler per round by ler}.

Finally, to quantify how decoding performance scales with code distance, we use the error suppression factor $\Lambda$ \cite{google2021exponential,google2023suppressing}, which characterizes the suppression of logical errors with increasing code distance. The logical error rate is expected to scale with distance as
\begin{equation}
\epsilon(d)=C\,\Lambda^{-\frac{d+1}{2}},\label{eq:lambda fit}
\end{equation}
where $C$ is a constant and $d$ is the code distance. A value of $\Lambda>1$ indicates that increasing the code distance suppresses the logical error rate, consistent with operation in the below-threshold regime. Larger values of $\Lambda$ correspond to faster suppression. When the noise is fixed, this corresponds to better decoding accuracy at large distances. When data is available for more than two code distances, $\Lambda$ can be estimated via a fit of $\log\epsilon(d)$ against $d$. When only two code distances $d$ and $d+2$ are available, this reduces to the simple pairwise ratio
\begin{equation}
\Lambda=\frac{\epsilon(d)}{\epsilon(d+2)},\label{eq:lambda 2 dist}
\end{equation}
which follows directly from the scaling relation above.

\subsection{Learning decoder priors from data}\label{subsec:learning_decoder_priors}

Decoder priors are often constructed from a representative theoretical noise model applied to a given circuit. Such priors are only an approximation of the true device noise. To improve them, it is desirable to estimate error probabilities directly from the device's own experimental detector data, so that the resulting priors more faithfully reflect the hardware. This requires a procedure for estimating the probability of an error mechanism from experimentally observed detector statistics, both for mechanisms already present in the DEM, and, as we show in Section \ref{sec:augmentation method}, for new candidate mechanisms considered during augmentation.

\subsubsection{The $p_{ij}$ method}\label{subsec:The-pij-method}

In this work, we estimate the probability of errors with the $p_{ij}$ method \cite{spitz2018adaptive,google2021exponential}, owing to its simplicity of implementation and computational efficiency. We therefore provide a brief overview of the method. Let $x_{i}$ and $x_{j}$ denote the binary detection outcomes of detectors $i$ and $j$ (equal to $1$ if the detector fires and $0$ otherwise), and let $p_{ij}$ denote the probability of an error mechanism that flips both detectors $i$ and $j$, while $p_{i}$ and $p_{j}$ denote the effective probabilities aggregating all error mechanisms that flip only detector $i$ or only detector $j$, respectively. The expected rate of observing simultaneous detector flips is then
\begin{equation}
    \left\langle x_{i}x_{j}\right\rangle =p_{ij}\left(1-p_{i}\right)\left(1-p_{j}\right)+\left(1-p_{ij}\right)p_{i}p_{j}.
\end{equation}
A joint detection event is observed either when the correlated error occurs and neither individual error occurs, or when both individual errors occur in the absence of the correlated error. Similar expressions can be derived for the individual detector firing rates $\left\langle x_{i}\right\rangle$ and $\left\langle x_{j}\right\rangle $:
\begin{align}
\left\langle x_{i}\right\rangle  & =p_{i}\left(1-p_{ij}\right)+\left(1-p_{i}\right)p_{ij},\\
\left\langle x_{j}\right\rangle  & =p_{j}\left(1-p_{ij}\right)+\left(1-p_{j}\right)p_{ij}.
\end{align}

By inverting these relationships, one can estimate the underlying error probabilities $p_{ij}$ from experimentally observed detector statistics $\left\langle x_{i}x_{j}\right\rangle $, $\left\langle x_{i}\right\rangle $ and $\left\langle x_{j}\right\rangle $: 
\begin{equation}
    p_{ij}=\frac{1}{2}-\frac{1}{2}\sqrt{1-\frac{4\left(\left\langle x_{i}x_{j}\right\rangle -\left\langle x_{i}\right\rangle \left\langle x_{j}\right\rangle \right)}{1-2\left\langle x_{i}\right\rangle -2\left\langle x_{j}\right\rangle +4\left\langle x_{i}x_{j}\right\rangle }}.
\end{equation}
Further details can be found in \cite{google2021exponential}. This framework has also been extended to higher-order correlated error mechanisms \cite{remm2026experimentally,blume2025estimating,takou2025estimating}, allowing us to estimate the probabilities of higher-order candidate mechanisms. Throughout this work, whenever we refer to the \emph{pairwise correlation} between two detectors, we mean the value $p_{ij}$ obtained from this inversion procedure, rather than the raw correlation between their firing outcomes.

\subsubsection{Recurrence edge method}\label{subsec:recurrence}

As the number of rounds executed in each experiment grows, so does the number of decoder priors that need to be estimated, making the learning process increasingly costly. However, instead of treating each hyperedge as an independent mechanism, we can take into consideration that many of the hyperedges are caused by the same physical error mechanisms that simply occur in different rounds~\cite{sivak2024optimization, bhardwaj2026adaptive, spitz2018adaptive}. Therefore, if the error rates of each such mechanism remain constant over time, we can identify the hyperedges that recur identically across the different rounds and assign each of them only a single, shared prior. We refer to this as the recurrence-edge method. This method removes the need to estimate each hyperedge separately, which substantially reduces the number of parameters to be learned, particularly in experiments with many rounds, and accelerates the learning process. Moreover, the observed detection events associated with each recurring instance can be pooled together to estimate this single probability. As a result, the different rounds of the experiment effectively act as additional samples when it comes to estimating these recurrent priors, thereby improving the accuracy of the estimated probabilities. The recurrence-edge method is only applicable when the physical error probabilities do not substantially drift over the duration of the experiment and when this condition holds, it enables both faster and more accurate estimation of the decoder priors for a given number of samples.

\section{Augmentation method}\label{sec:augmentation method}

We wish to improve the decoder by giving it a more physically accurate picture of the noise, and in particular, by making it aware of correlated errors. One could attempt this by explicitly adding correlated noise terms to the theoretical model, but this would require assuming which pairs of qubits are affected and the structure of the correlations. Since correlated noise is highly sensitive to the specific calibration and fabrication details of a device, this approach is not practical. Instead, we take a data-driven route: we construct a method to augment the decoder priors by discovering entirely new error mechanisms directly from experimental detector-firing statistics.

Our goals for the augmentation method are threefold. First, the method should be general, working for any code and any DEM without being tailored to a specific architecture. Second, it should be efficient, keeping the computational cost practical. Third, it must preserve decoder compatibility:  we require the augmented DEM to remain compatible with any decoder that can operate on the original DEM. To be compatible with decoding at all, every newly added error mechanism must first be assigned a logical effect. Beyond this, each newly discovered hyperedge must be accompanied by a valid decomposition into pre-existing elements of the original DEM. Enforcing a valid decomposition ensures that any decoder applicable to the original DEM remains applicable to the augmented DEM. In particular, matching-based decoders remain applicable whenever they were applicable to the original DEM.

\begin{figure}
\centering
\includegraphics[width=\textwidth]{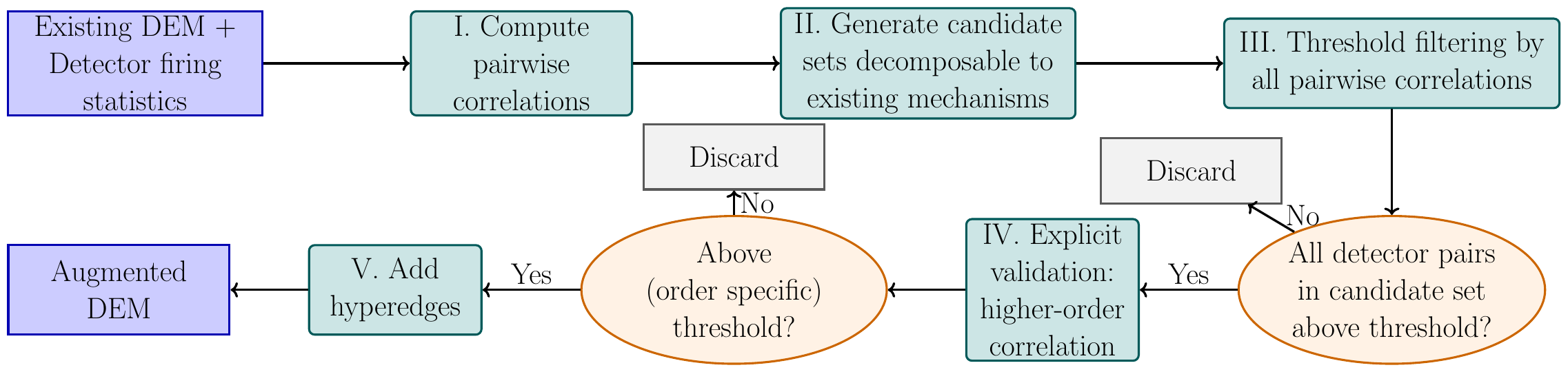}\caption{A high level overview of the DEM augmentation procedure. The purple boxes denote the inputs and outputs of the augmentation method: we input an existing DEM and experimental detector statistics and produce a new, augmented DEM used to calibrate a correlation-aware decoder. Green boxes denote the five main steps of the pipeline, each described in detail in the following subsections: computing pairwise correlations (Subsection \ref{subsec:aug-step-1-pairwise-computing}), generating candidate mechanisms (Subsection \ref{subsec:aug-step-2-candidates}), pairwise cutoff filtering (Subsection \ref{subsec:aug-step-3-pairwise-filtering}), explicit higher-order validation (Subsection \ref{subsec:aug-step-4-higher-order}), and adding validated mechanisms to the DEM (Subsection \ref{subsec:aug-step-5-add-hyperedges}). Orange circles denote the two filtering decisions at which candidates may be discarded: failing the pairwise cutoff filter, or failing explicit higher-order validation.}\label{fig:flow chart}
\end{figure}

We now present a method that satisfies all of these requirements. The method does not require code-specific assumptions and should extend naturally to any DEM, but we will use the surface code \cite{kitaev2003fault,fowler2012surface} as a running example for concreteness and clarity. Figure \ref{fig:flow chart} shows a high-level overview of the full augmentation pipeline, which we summarize here before describing each step in detail in the following subsections. The method is carried out in five steps. First, we compute the pairwise correlation between all pairs of detectors using the experimental data (for example, using the $p_{ij}$ method discussed in Section \ref{subsec:The-pij-method}). Second, we generate candidate higher-order mechanisms by combining pairs of existing mechanisms in the DEM. In step three, we use the pairwise correlations computed in step one as a fast pre-filter, discarding candidates whose constituent detector pairs are not all sufficiently correlated. In step four, surviving candidates are explicitly validated by directly computing their higher-order correlation. Once again, if a candidate does not meet a correlation cutoff, it is discarded. Finally, in step five, validated candidates are assigned a logical effect and a decomposition inherited from their constituent mechanisms, and are added to the DEM as new hyperedges.

\subsection{Step one: compute all pairwise correlations}\label{subsec:aug-step-1-pairwise-computing}

We begin with the first step of the augmentation procedure shown in Figure \ref{fig:flow chart}: computing, from the detector data, pairwise correlations between all pairs of detectors in the circuit. 
Any method capable of computing correlations may be used for this purpose. We adopt the $p_{ij}$ method described in Section \ref{subsec:The-pij-method}. We note that a naive approach for finding new error mechanisms would be to compute all higher-order correlations up to a maximum order, and then search for correlations that are not already explained by mechanisms in the input DEM. This quickly becomes intractable as the number of possible detector subsets grows exponentially with order, making an exhaustive search over all higher-order candidates computationally infeasible. We restrict to computing pairwise correlations, where the number of pairs grows quadratically with the number of detectors, to ensure the method remains practical to implement.

This choice is motivated by a key insight: any higher-order correlated error mechanism necessarily manifests as a pairwise correlation between every pair of detectors it involves. For instance, if a single error mechanism triggers detectors $A$, $B$ and $C$ simultaneously, then $AB$, $AC$, and $BC$ will all be pairwise correlated, even though no pairwise mechanism alone explains the full event $ABC$. Consequently, while pairwise correlations alone are not sufficient to confirm the existence of a higher-order mechanism (a point to which we return in Step three, Subsection \ref{subsec:aug-step-3-pairwise-filtering}), their presence is a necessary condition. This lets us use the computed pairwise correlations as a fast, necessary-condition screening tool: rather than exhaustively testing every higher-order detector subset, we can restrict attention to those subsets whose pairwise correlations have already been confirmed, drastically narrowing the candidate space when performing more rigorous validation in later steps.

\subsection{Step two: generate candidate sets}\label{subsec:aug-step-2-candidates}

In the second step of the procedure, we construct sets of candidate correlated error mechanisms by combining pairs of preexisting mechanisms in the input DEM. While in this work we only focus on pairwise candidates, we note that the method generalizes naturally to combinations of any number of existing mechanisms. 
By constructing candidates using pre-existing mechanisms as building blocks, we observe several key advantages:

\begin{itemize}
\item Captures correlations between any pair of existing error mechanisms: 
As our method generates candidates by combining all pairs of error mechanisms in the input DEM, we are able to uncover correlations between any pair of existing noise mechanisms. Using our method, we may capture space-like, time-like and combined space-time correlations throughout the circuit.

\item Limits the candidate space: Since candidates are restricted to combinations of existing mechanisms, only a small fraction of all possible detector subsets qualify. Most subsets of detectors will not admit a decomposition into existing mechanisms. This smaller candidate space improves computational efficiency and, more importantly, statistical robustness: evaluating fewer candidates lowers the likelihood of flagging spurious correlations caused by finite-sample noise.

\item Provides natural decompositions and assignment of logical observables: Since every candidate is constructed from known mechanisms, it inherits a natural decomposition and a well-defined logical effect. Both are essential for ensuring that the augmentation does not impose additional restrictions on the decoder and, in particular, that compatibility with correlated graph-based decoders is preserved when augmenting graph-compatible input DEMs.
\end{itemize}

\begin{figure}
\centering
\includegraphics{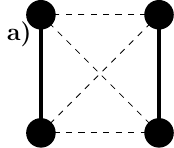}\hspace{25pt}\includegraphics{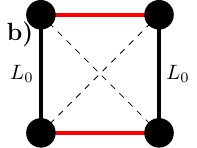}\hspace{15pt}\includegraphics{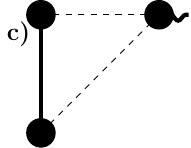}\hspace{5pt}\includegraphics{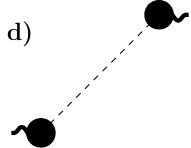}\hspace{10pt}\includegraphics{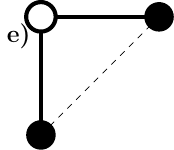}
\caption{Examples of candidate error mechanisms obtained by pairing two existing mechanisms, illustrated for codes where the basic error mechanisms are edges and singletons. Vertices represent detectors, solid lines represent edges present in the original DEM, dashed lines represent detector connections not present in the original DEM, and curly lines indicate singletons (connected to a boundary). A logical label near an edge indicates that the corresponding error mechanism also flips that logical observable. (a) An order-4 candidate formed from two non-overlapping existing edges. (b) An order-4 candidate that admits two distinct decompositions into existing mechanisms. The first decomposition is highlighted by the black vertical edges. Since each edge in the decomposition carries the logical effect of flipping $L_0$, their XOR yields no logical flip. The second decomposition is shown by the two red horizontal edges, which also yields no logical flip. As both decompositions assign the same logical effect, the candidate is valid. (c) An order-3 candidate composed of an existing edge and an existing singleton. (d) An edge candidate composed of two existing singletons, denoted a decomposed edge. (e) An edge candidate, denoted a long-range edge, composed of two existing edges that share exactly one detector. The empty circle marks this shared detector, which does not trigger when both constituent error mechanisms occur together, and so is not part of the candidate. The resulting candidate is an edge connecting the two remaining detectors.}\label{fig:candidates}
\end{figure}

We can additionally impose a cutoff on the order of the candidates, i.e. the maximum number of detectors a candidate may involve, allowing us to control the complexity of the error mechanisms we search for. To build intuition, here we present the candidate set for the surface code. We assume that the input DEM has already been decomposed into a graph-like model (see Subsection \ref{subsec:correlated_decoding}), ensuring that the existing mechanisms are edges and singletons, rather than higher-order hyperedges. Figure \ref{fig:candidates} shows the possible sets of candidate error mechanisms constructed when edges and singletons form building blocks. The different types of candidate mechanisms are:

\begin{itemize}
\item Order-4 hyperedges: formed by pairing any two non-overlapping existing edges. See Figures \ref{fig:candidates}(a) and \ref{fig:candidates}(b).

\item Order-3 hyperedges: formed by pairing any existing edge and singleton that do not overlap. See Figure \ref{fig:candidates}(c).

\item Decomposed edges: formed by pairing any two existing singletons. See Figure \ref{fig:candidates}(d).

\item Long-range edges: formed from two existing edges that share exactly
one detector. When both error mechanisms occur simultaneously, the
shared detector is flipped twice and cancels, leaving a longer-range
edge between the two remaining detectors. See Figure \ref{fig:candidates}(e).
\end{itemize}

Note that we deliberately exclude singleton candidates, as they are too susceptible to statistical noise, even though correlated error mechanisms can give rise to them.

\subsection{Step three: pairwise cutoff filtering}\label{subsec:aug-step-3-pairwise-filtering}

Given the set of candidate mechanisms obtained in Subsection \ref{subsec:aug-step-2-candidates}, we now need to filter the candidates and retain only those that show correlated behavior. As noted in Subsection \ref{subsec:aug-step-1-pairwise-computing}, any genuine higher-order correlation will produce pairwise correlations between every pair of detectors contained within the candidate. We exploit this as a fast pre-filter. For each candidate we check whether all pairwise correlations among its detectors exceed a predefined cutoff. Candidates that fail this check are discarded before reaching the more expensive explicit validation step detailed in Subsection \ref{subsec:aug-step-4-higher-order}.

The choice of cutoff involves a trade-off. A high cutoff is conservative and may miss genuine correlated error mechanisms. A low cutoff admits more candidates but also increases the risk of including spurious candidates driven by statistical fluctuations. Rather than choosing the cutoff based on statistical significance alone, we instead treat it as a free parameter. We run the full augmentation procedure with several candidate cutoff values, decode with each of the resulting augmented DEMs, and select the cutoff that yields the lowest logical error rate. To avoid overfitting, we use separate datasets for determining these cutoffs in the augmentation procedure. Note that the cutoff also affects the runtime of the augmentation method itself, too low a cutoff admits too many candidates, and as the main runtime cost of the algorithm is in computing the higher order correlation, this makes the procedure computationally expensive. In practice, the search is limited to values for which the augmentation runs in a reasonable time.

\subsection{Step four: explicit validation via higher order correlation} \label{subsec:aug-step-4-higher-order}

We now apply a second filter to the reduced set of candidates that passed step three. Passing the pairwise filter is a necessary but not sufficient condition for a candidate to represent a genuine higher-order mechanism. Pairwise correlations can appear between all detectors in a set without a direct higher-order correlation among all detectors in the candidate. We therefore perform an explicit validation step, directly computing the higher-order correlation for each remaining candidate (for example, using the higher-order generalizations of the $p_{ij}$ method \cite{remm2026experimentally,blume2025estimating,takou2025estimating}), and retaining only those that exceed a predetermined cutoff. While this explicit validation step is more computationally expensive, the pairwise filtering in step three drastically reduces the number of candidates, making this explicit computation entirely tractable. We may also apply different cutoffs for candidates of different orders. The probability of observing a spurious higher-order correlation due to statistical noise decreases as the order of the candidate increases, so higher-order candidates can afford to be held to a lower cutoff.

A separate subtlety arises when explicitly computing the higher-order correlation for a candidate. The observed correlation between a set of detectors $\{A_1, A_2, \dots, A_n\}$ can be influenced by any error mechanism that includes these detectors, even one that involves additional detectors. For example, if detectors $A_1$ and $A_2$ are both involved in a genuine order-3 mechanism together with a third detector $B$, the order-3 mechanism will itself contribute to the observed pairwise correlation $A_1 A_2$, inflating any naive estimate of a distinct order-2 (edge) candidate between $A_1$ and $A_2$ alone.

To accurately estimate the probability of an error mechanism involving exactly $\{A_1, A_2,\dots, A_n\}$ we must remove contributions from all higher-order mechanisms that also include these detectors. We address this by validating candidates in decreasing order, starting with the largest order candidate present. By the time we analyze a given candidate all higher-order mechanisms sharing its detectors have already been confirmed, and their contributions can be accounted for using one of the two methods below:

\begin{itemize}
\item Data Erasure Method:
For each confirmed higher-order mechanism, we remove its contribution directly from the raw data before analyzing lower-order candidates. Specifically, in any shot where all detectors of a confirmed higher-order mechanism fire simultaneously, we reset those detector values to zero, effectively treating that shot as if the higher-order event had not occurred. Repeating this for every confirmed higher-order mechanism yields a dataset in which lower-order candidates can be analyzed as if the higher-order mechanisms were absent.

\item Analytical Subtraction Method: 
Alternatively, we can leave the data untouched and instead calculate analytically how much each confirmed higher-order mechanism is expected to contribute to the observed correlation of the lower-order candidate, using the higher-order mechanism's own previously assigned probability. Subtracting this expected contribution from the observed correlation isolates the portion attributable to the candidate mechanism alone, from which its probability can then be estimated (analogous to the inversion used in the $p_{ij}$ method in Subsection \ref{subsec:The-pij-method}).
\end{itemize}

Each method has trade-offs. The analytical method can propagate errors in the estimated probabilities of higher-order mechanisms down to lower-order ones. The data erasure method avoids this issue, as it directly removes events from the data, but it can be inaccurate if events overlap (i.e., not all detectors of a higher-order mechanism fire due to interference from other mechanisms). The choice between these methods depends on factors such as the accuracy of assigned probabilities, the number of samples, and the device’s error rates. We used the data erasure method for the simulated results and the analytical subtraction method for the experimental results, see Subsection \ref{subsec: results simulated}.

\subsection{Step five: add the new error mechanism}\label{subsec:aug-step-5-add-hyperedges}

In the final step of the procedure, we add validated candidates to the DEM. To do so, we first determine each candidate's logical effect. By construction, each candidate can be decomposed into existing mechanisms, each with well-defined observables effects. The logical effect of the candidate is obtained by XORing the logical effects of its constituents. In some cases, as illustrated in Figure~\ref{fig:candidates}(b), a candidate may have multiple valid decompositions. We briefly check that all decompositions yield the same observable. If not, the candidate is discarded as it is not physically meaningful \footnote{For codes with sufficiently large distance relative to the number of correlated mechanisms considered, this consistency is always satisfied and the check serves mainly as a sanity check. For small-distance codes, however, correlated mechanisms can wrap around the code in non-trivial ways and produce inconsistencies, so this step becomes a meaningful filter.}. Candidates that pass this check are added as new hyperedge mechanisms to the DEM. The method supports both decomposed and undecomposed DEMs, depending on the decoder to be used:

\begin{itemize}
\item Undecomposed DEM: The new instruction lists all detectors in the candidate and the (consistent) logical effect.
\item Decomposed DEM: The new instruction uses a separator \cite{gidney2021stim} to split the hyperedge into its constituent components, each annotated with its logical effect. If multiple decompositions exist, we add a separate instruction for each decomposition, dividing the probability equally among them as a first-order approximation.
\end{itemize}

We include all possible decompositions for two main reasons. First, considering the two-pass decoder, we want to ensure that correlations are captured regardless of which subset of the hyperedge is identified during the initial decoding pass. As discussed in Subsection \ref{subsec:correlated_decoding} and Appendix \ref{app:adaptivity map}, correlated decoding operates in two stages, with probabilities updated based on the edges selected as part of the recovery operator in the first pass. Therefore, the correlation of a hyperedge is only considered if the first pass identifies one of its constituent edges. By including all decompositions, we increase the chances that the decoder will select an edge in the first pass that triggers the consideration of the hyperedge’s correlation. Second, since the correlations between detector firings are unaffected by how the hyperedge is decomposed, there is no principled basis for favoring one decomposition over another. We therefore maintain symmetry by treating all decompositions equally.

Each newly discovered mechanism is assigned the probability estimated in Section \ref{subsec:aug-step-4-higher-order} using the explicit higher-order correlation formula. The purpose of the augmentation procedure is to discover new error mechanisms, not to refine the probabilities of existing ones. The probabilities of mechanisms already present in the DEM remain unchanged, even though they may participate in correlated error processes. Probability refinement can instead be performed as a separate step after augmentation, for example using the $p_{ij}$ method \cite{remm2026experimentally,blume2025estimating,takou2025estimating} or a machine-learning-based approach \cite{sivak2024optimization}. In Section \ref{sec:results}, we present results both with augmentation alone and, where specified, with an additional $p_{ij}$-based probability-learning step applied to all mechanisms in the augmented DEM.

\begin{figure}
\centering
\includegraphics{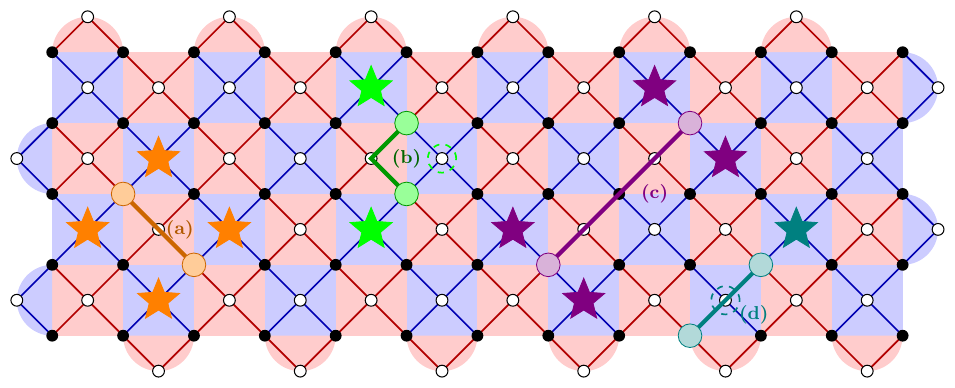}\caption{ A visualization of the surface code. Data qubits are black, ancilla qubits are white, $X$-type stabilizers are red, and $Z$-type stabilizers are blue. A correlated error mechanism applies a joint $XX$ error between two data qubits. Affected qubits are marked by a filled circle, and the detectors that fire due to the correlated error mechanism are marked by stars of the same color. If there is a detector that does not fire, since two affected qubits from the same error mechanism trigger it, the qubit is marked by a dashed circle of the same color. (a) An error mechanism of correlation distance 2, that leads to a new order-4 hyperedge, orange stars. (b) An error mechanism of correlation distance 2 that leads to a new long range edge, green stars. (c) An error mechanism of correlation distance 4 that leads to a new hyperedge of order 4, magenta stars. (d) An error mechanism of correlation distance 2 that leads to a new singleton, teal stars.}\label{fig:crosstalk types}
\end{figure}

\section{Results}\label{sec:results}

We now apply our DEM augmentation procedure to simulated and experimental surface code quantum memory experiments. For simulations, we apply SI1000 noise (Appendix \ref{app:si1000}) as a baseline and supplement it with additional correlated noise. For testing the augmentation procedure against experimental data, we use data generated from Google's below-threshold experiments \cite{google2024zenodo}. All results are decoded using a two-pass correlated MWPM decoder \cite{edmonds1965paths,dennis2002topological,fowler2012surface}, implemented in PyMatching \cite{higgott2022pymatching,higgott2025sparse}.

\subsection{Validation on simulated data}\label{subsec: results simulated}

We first evaluate the augmentation method on simulated circuits. Our starting point is a surface code simulation using SI1000 noise (Appendix \ref{app:si1000}) with noise parameter $p=10^{-3}$. We supplement the base SI1000 model with correlated errors loosely inspired by the crosstalk characterized in Google's device \cite{google2023suppressing,google2021exponential}. 

To apply the correlated noise, we define a \emph{qubit interaction graph} in which two qubits are connected by an edge if they share a coupler enabling a two-qubit gate. The \emph{correlation distance} of a pair of qubits is then the graph distance (length of the shortest path) between them in the interaction graph. For example, qubits connected by a single coupler have correlation distance 1, two qubits that do not interact directly but share a common neighbor (i.e. each qubit may interact with a common third qubit) have correlation distance 2. This generalizes naturally to any correlation distance $n$. We model an additional correlated error mechanism as a joint Pauli $XX$ term applied to data qubits at a fixed correlation distance, activated when both qubits participate in separate two-qubit gates within the same circuit layer. This is designed to represent unwanted interactions between qubits that are actively participating in a two-qubit gate layer \cite{arms2025estimating}. Representative qubit pairs of correlation distance 2 are illustrated in Figure \ref{fig:crosstalk types}, labeled (a) and (b), and of correlation distance 4, labeled (c). The correlation noise parameter used throughout is $p_{\mathrm{corr}}=3\times10^{-4}$.

To reflect an experimenter who is unaware of correlated noise present in the device, we proceed as follows. We sample a set of detector events from the circuit containing correlated noise and split the data into two disjoint sets: one for DEM augmentation and one for decoding. Along with the augmentation subset, we submit a \emph{vanilla DEM}, generated using a Stim circuit with standard SI1000 noise and no additional correlated noise ($p_{\mathrm{corr}} = 0$), to the augmentation procedure. This produces the \emph{augmented DEM}, which is used to decode the disjoint set of detector events reserved for decoding. Since the simulation data originates from a circuit with known correlated noise, we also have access to the \emph{oracle DEM}, generated directly from the Stim circuit including correlated noise. The oracle DEM serves as a ground truth and allows us to evaluate the augmentation method in several complementary ways. We first check whether the augmentation method succeeds in recovering most of the error mechanisms absent from the vanilla DEM, but present in the oracle DEM, while introducing no (or few) spurious mechanisms. Secondly, we test whether using the augmented DEM yields improved decoder accuracy relative to the vanilla DEM, approaching the values obtained with the oracle DEM. 

As suggested in Subsection \ref{subsec:aug-step-4-higher-order}, we choose different cutoffs for candidates of different order during the augmentation procedure. We define $c_n$ as the correlation cutoff used for candidates of order $n$. Unless stated otherwise, we use the following cutoff values: $c_4=10^{-4}$, $c_3=1.5\times10^{-4}$, and $c_2=2\times10^{-4}$. These values were found to strike a good balance between decoder accuracy and augmentation efficiency, see Appendix \ref{app:Simulated threshold}. Low cutoffs were found to be more beneficial for decoding, but the choice of parameters here is not so low as to make the candidate generation prohibitively slow (recall that lower cutoffs admit more candidates at each filtering stage, increasing runtime). We limit the maximum order of candidates to four. We use the $p_{ij}$ method to learn the probabilities of the new mechanisms.

For computational efficiency, all simulations use 10 rounds of error correction. Although the code distances we consider exceed 10, this is not problematic for quantum memory experiments. Since the logical state is initialized as an eigenstate of the logical operator, error strings cannot wrap non-trivially around the code through the time boundaries \cite{fowler2012surface,dennis2002topological}, so the number of rounds does not need to scale with the code distance. However, using too few rounds degrades the quality of the $\Lambda$ fit (Subsection \ref{subsec:logical error rate and lambda}). We therefore chose 10 rounds, which yields a robust linear fit in the $\Lambda$ plot, as shown in Figure \ref{fig:lambda plot}.

\subsubsection{Recovery of missing error mechanisms.}\label{subsec:recovery}

\begin{figure}
\centering
\includegraphics[scale=0.3]{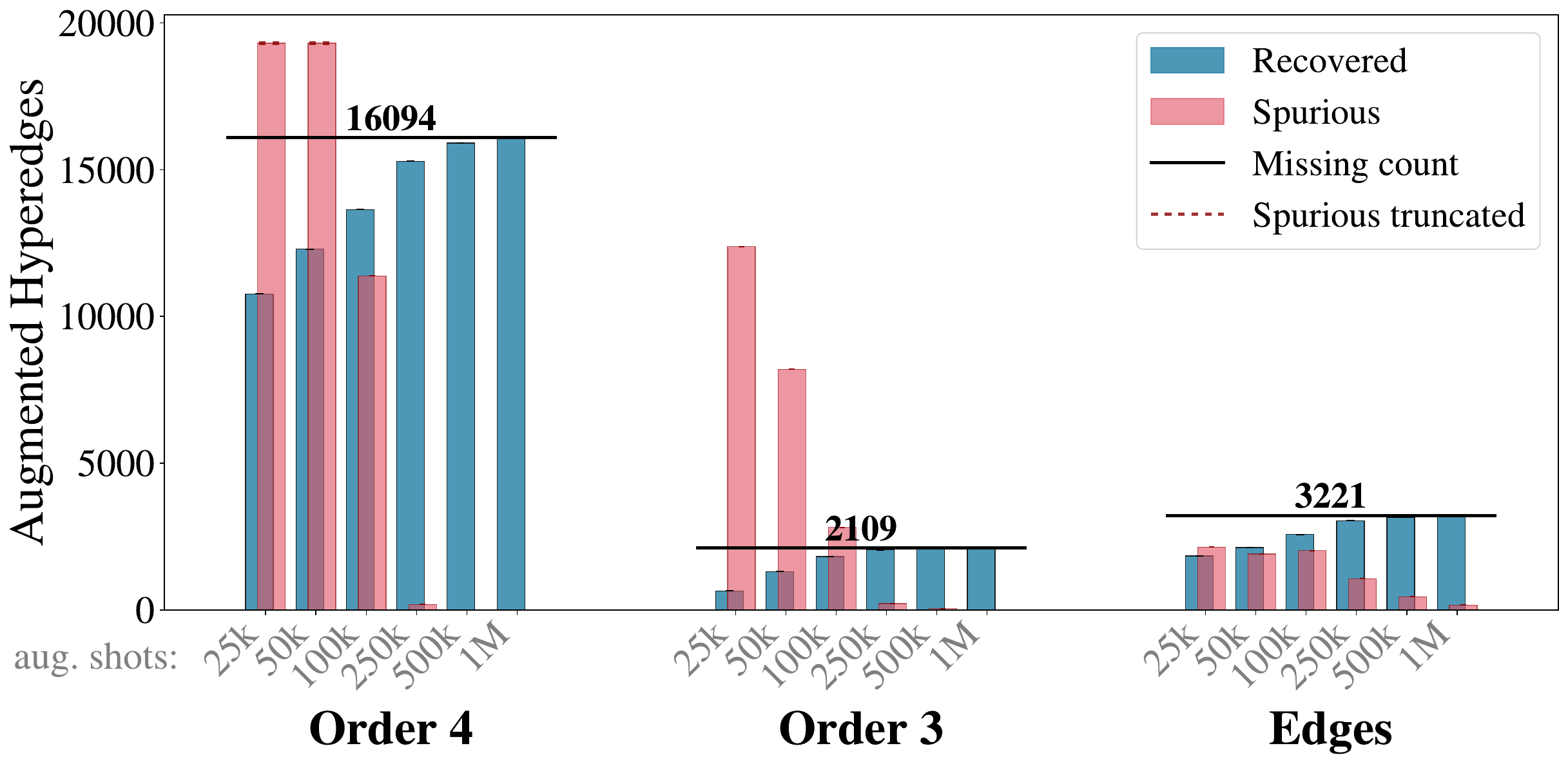}\caption{Hyperedge recovery performance of the augmentation method for a distance-15 surface code with correlation distance 2 and strength $p_{corr}=3\times10^{-4}$. The x-axis is grouped by error mechanism type,  order-4 hyperedges, order-3 hyperedges, and edges (both decomposed edges and long range edges), corresponding to error mechanisms present in the oracle DEM but absent from the vanilla SI1000 DEM. Within each group, sub-ticks denote the number of samples used to train the augmentation method. Solid horizontal black lines indicate the true count of missing mechanisms per type from the vanilla DEM. Blue bars show the number of correctly recovered mechanisms, while red bars show spurious mechanisms erroneously introduced by the augmentation procedure. Spurious mechanism counts exceeding the axis range are truncated, for order-4 hyperedges their values are 263,174 and 76,983 for 25k and 50k shots, respectively. As the number of samples increases, the method converges cleanly. Spurious mechanisms are rapidly suppressed to negligible levels, while nearly all missing mechanisms are correctly recovered.}
\label{fig:recovery plot}
\end{figure}

We begin by analyzing how well the augmentation procedure recovers the correlated error mechanisms present in the oracle DEM. Figure \ref{fig:recovery plot} shows the number of recovered and spurious error mechanisms as a function of the number of augmentation samples, for code distance 15 and noise with correlation distance 2. As can be seen, the augmentation method converges cleanly. As the number of augmentation samples increases, spurious mechanisms become negligible and virtually all true missing mechanisms are recovered, confirming that the method behaves as expected and providing a practical estimate of the sample budget required for convergence for this error probability. We note that this analysis was performed without the recurrence edge method discussed in Subsection \ref{subsec:recurrence}, which exploits additional rounds as a substitute for additional samples, allowing convergence with fewer samples when more rounds are available. Therefore, Figure~\ref{fig:recovery plot} should be taken as evidence for convergence rather than quantitative cost of converging. Similar qualitative behavior was observed for simulations with correlation distance 4 noise mechanisms, though with a somewhat larger number of spurious mechanisms owing to the larger initial candidate pool.

We also verify that the pairwise correlation filter significantly reduces the number of candidates for which the higher-order correlation must be computed in step four of the augmentation procedure. In this example, the pairwise correlation filter reduced the candidate set to just $0.3\%$ (from 62M to 194k) and $0.5\%$ (from 6M to 31k) of its original size for order-4 and order-3 hyperedges, respectively, leaving only this small fraction for the second, more time-consuming filter to process, though these numbers vary with the code distance and depend on whether the recurrence edge method is used (Subsection \ref{subsec:recurrence}).

\subsubsection{Decoding performance.}

Given that the augmentation procedure correctly recovers correlated error mechanisms present in the oracle DEM, we now turn our attention to decoder accuracy. We assess accuracy using the error suppression parameter, $\Lambda$, defined in Subsection \ref{subsec:logical error rate and lambda}.  Higher $\Lambda$ values indicate more effective error suppression as the code grows due to improved decoder accuracy. We use one million samples for DEM augmentation and four million samples for decoding. An analysis of how the number of augmentation samples affects the logical error rate (LER) is provided in Appendix~\ref{app:ler_samples}.

Figures \ref{fig:lambda plot}(a) and \ref{fig:lambda plot}(b) show LER plots and fitted $\Lambda$ values for simulations containing correlation distance 4 and 2 noise, respectively. For correlation distance 4, panel (a), the augmented DEM achieves $\Lambda=2.06 \pm 0.01$, demonstrating a $20\%$ improvement over the $\Lambda=1.71 \pm 0.01$ obtained when decoding using the vanilla DEM. Furthermore, the augmented DEM performs nearly as well as the oracle DEM, which achieved $\Lambda=2.08 \pm 0.01$. At code distance 15, the largest considered, the LER is $(2.74 \pm 0.04) \times 10^{-5}$ for the oracle DEM, $(2.83 \pm 0.04) \times 10^{-5}$ for the augmented DEM, and $(1.14 \pm 0.08) \times 10^{-4}$ for the vanilla DEM. Both the oracle and augmented DEMs outperform the vanilla DEM by more than a factor of 4, underscoring the importance of accurately modeling correlated error mechanisms.

For correlation distance 2, shown in panel (b), the augmented DEM achieves a smaller, though still significant, improvement over the vanilla baseline, and does not fully close the gap with the oracle DEM. To attempt to close this gap, following the discussion in Appendix \ref{app:residual discrepancy}, we also re-learned the weights of all error mechanisms in the augmented DEM via the $p_{ij}$ method after augmentation. The augmented DEM obtained $\Lambda=2.82 \pm 0.03$, demonstrating a $15\%$ improvement over the $\Lambda = 2.46 \pm 0.02$ obtained for the vanilla DEM. The oracle DEM in this case achieves $\Lambda=3.02 \pm 0.04$. For distance 15, the LER is $(3.16 \pm 0.13) \times 10^{-6}$ for the oracle DEM, $(4.63 \pm 0.15) \times 10^{-6}$ for the augmented DEM, and $(1.54 \pm 0.03) \times 10^{-5}$ for the vanilla DEM. We find the augmented DEM reduces the LER by more than a factor of 3 compared to the vanilla DEM.

The reason the accuracy gap remains unresolved at correlation distance 2, while closing completely at correlation distance 4, is addressed in detail in Appendix~\ref{app:residual discrepancy}. In brief, at correlation distance 2, multiple valid decompositions of new hyperedges can lead to differences in probability assignment of the learned mechanisms compared to the oracle DEM. These factors are less significant at correlation distance 4.

\begin{figure}
\centering
\includegraphics[scale=0.41]{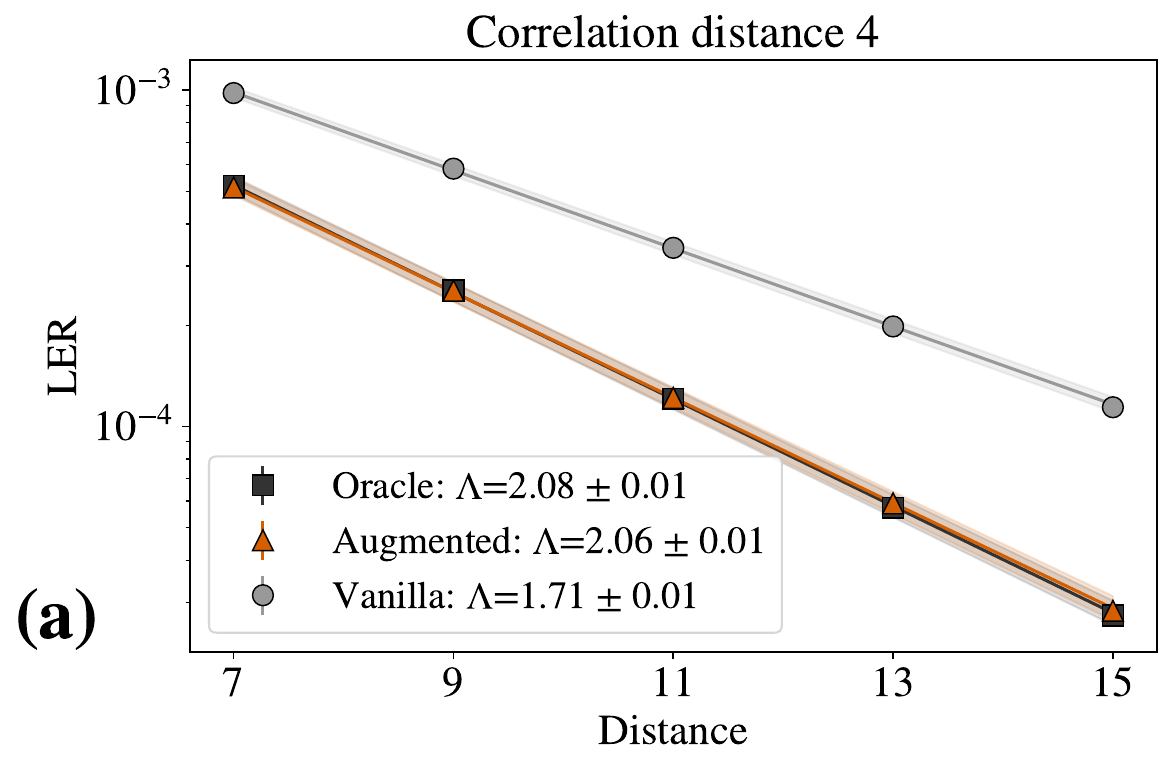}
\hspace{0.8cm}
\includegraphics[scale=0.41]{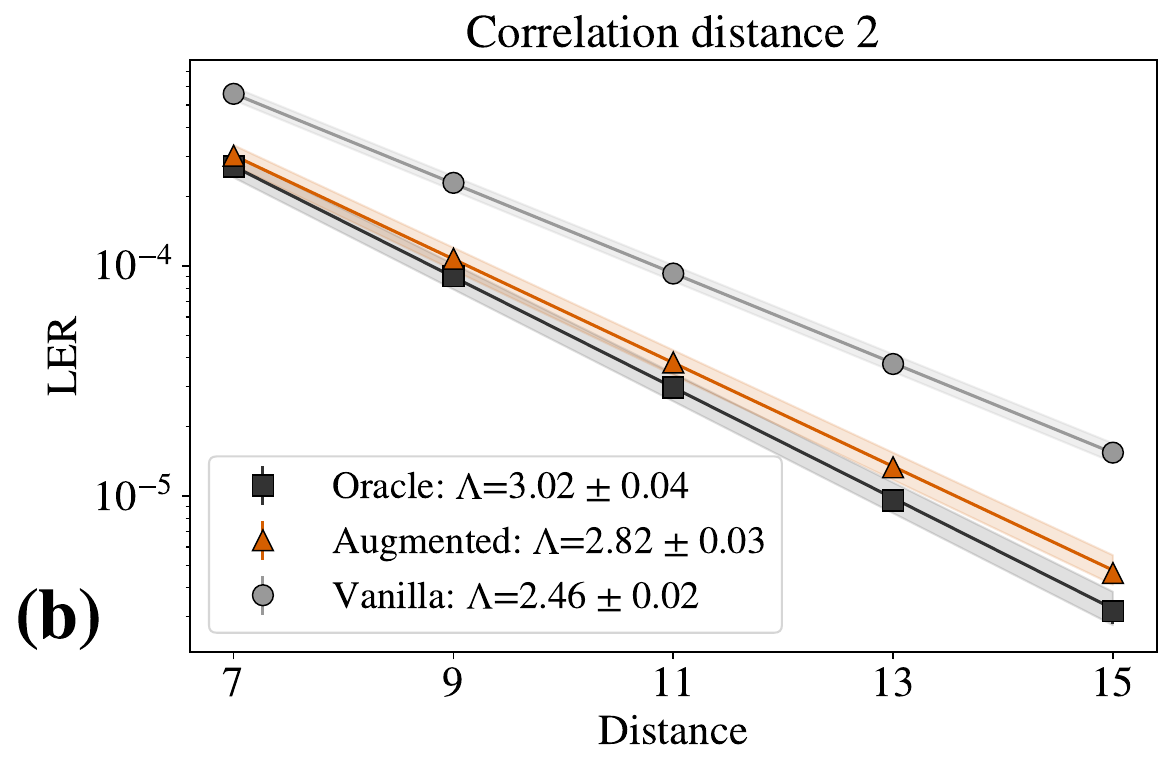}
\caption{$\Lambda$ plots for simulated experiments with code distances 7-15. A million samples were used for augmentation and another 4 million were used for decoding. (a) Correlation distance 4. The augmented DEM (orange triangles) has $\Lambda=2.06 \pm 0.01$ and so almost fully succeeds in recovering the $\Lambda=2.08 \pm 0.01$ of the oracle DEM (black squares), which is a significant improvement compared to the $\Lambda=1.71 \pm 0.01$ of the vanilla DEM (gray circles). (b) Correlation distance 2. Following the discussion in Appendix \ref{app:residual discrepancy} this time we also re-learned the weights of all error mechanisms in the augmented DEM, via the $p_{ij}$ method. The augmented DEM has a $\Lambda=2.82 \pm 0.03$, which does not fully recover the $\Lambda=3.02 \pm 0.04$ of the oracle DEM, it is still a significant improvement over the vanilla $\Lambda=2.46 \pm 0.02$.}\label{fig:lambda plot}
\end{figure}

\subsection{Decoding on Experimental Data}\label{subsec:decoding experimental}

\begin{figure}
\centering
\includegraphics[scale=0.4]{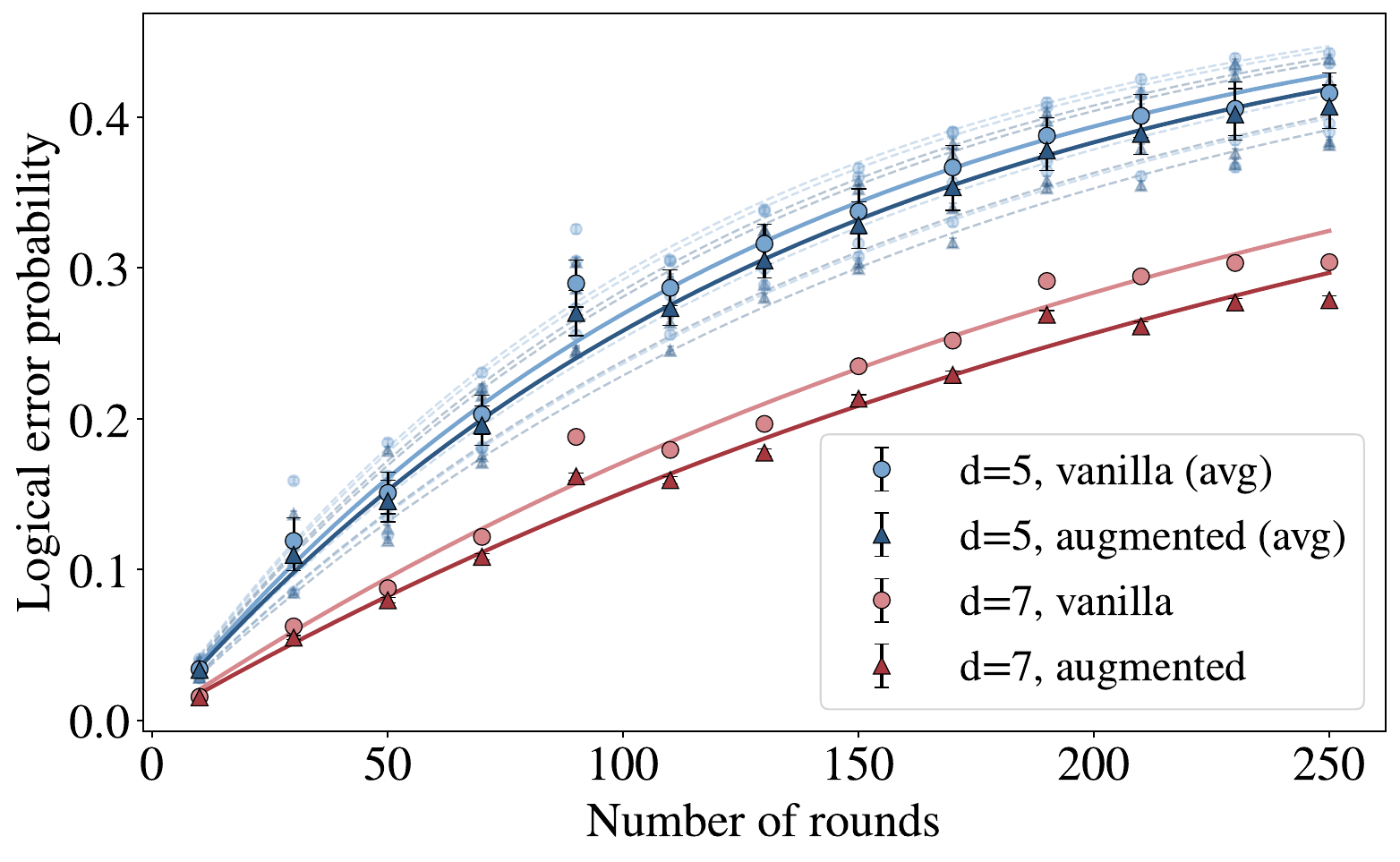}\caption{Logical error probability vs.\ number of rounds, decoded with the vanilla RL-optimized DEM (light color, circles) and the augmented DEM (dark color, triangles), on Google's experimental data \cite{google2024zenodo}. Blue: distance $d=5$ experiments showing four individual circuits with semi-transparent markers. Opaque markers denote the average LER over all circuits. Red: distance $d=7$ experiment. 
Solid curves show fits to Eq.~\eqref{eq:ler by ler per round}, yielding an LER of $(3.86\pm0.11)\times10^{-3}$ (vanilla) vs.\ $(3.62\pm0.10)\times10^{-3}$ (augmented) at $d=5$, and $(2.09\pm0.01)\times10^{-3}$ (vanilla) vs.\ $(1.80\pm0.01)\times10^{-3}$ (augmented) at $d=7$. The corresponding error suppression factors, calculated using Eq.~\eqref{eq:lambda 2 dist}, are $\Lambda=1.85\pm0.05$ (vanilla) and $\Lambda=2.02\pm0.06$ (augmented).}
\label{fig:experimental decoding}
\end{figure}

We now turn to evaluating the augmentation method on real experimental data, using publicly available measurements from Google's superconducting device \cite{google2024zenodo}. A key difference compared to the simulated setting is the absence of a ground truth: we can therefore only report the LER of the augmented DEM relative to that of the vanilla DEM. For the vanilla DEM we use the reinforcement learning (RL)-optimized DEM weights supplied with the experimental data \cite{google2024quantum}. We construct the augmented DEM by applying the augmentation method to this vanilla DEM, assigning weights to all mechanisms, both pre-existing and newly added, using the $p_{ij}$ method after augmentation. Importantly, the augmented DEM is at an inherent disadvantage in this comparison as the $p_{ij}$ method is strictly less powerful than RL optimization. An additional post-processing step, relearning weights for the augmented DEM via RL, could further improve decoding accuracy, as could improvements to how the conditional reweighting is done in the correlated decoding \cite{de2026correlated}, though both directions are outside the scope of the current work. Despite this, Figure~\ref{fig:experimental decoding} shows that augmentation still yields a measurable reduction in logical error rate relative to the RL-optimized vanilla baseline, offering further evidence that newly discovered mechanisms represent genuine physical error sources.

Since only 50k samples are available, we cannot study convergence across sample budgets as we did in the simulations shown in Figure \ref{fig:decoding plot}. However, measurements are available across many different numbers of rounds, each with 50k samples. We leverage this by applying the recurrence edge method described in Subsection~\ref{subsec:recurrence}, effectively obtaining more samples for the augmentation procedure. We use the cutoff values $c_4=5\times10^{-5}$, $c_3=\times10^{-4}$, and $c_2=2\times10^{-4}$, slightly lower than those used for the simulated data. The smaller code distances here allow for lower cutoffs while keeping computation time reasonable. In Figure~\ref{fig:experimental decoding} we show the logical error probability as a function of the number of rounds for both the vanilla RL-optimized DEM and the augmented DEM for distances 5 and 7. Since four distinct circuits are available at distance 5, we additionally report their average per-round values. Fitting this data according to Eq.~\eqref{eq:ler by ler per round}, we extract the LER for each distance.

For distance 7, the augmented DEM achieves an LER of $(1.80 \pm 0.01) \times 10^{-3}$, improving upon the vanilla RL-optimized DEM's $(2.09 \pm 0.01) \times 10^{-3}$ by $14\%$. For distance 5, the augmented DEM achieves an average LER of $(3.62 \pm 0.10) \times 10^{-3}$, improving upon the vanilla RL-optimized DEM's $(3.86 \pm 0.11) \times 10^{-3}$ by $6\%$. Using Eq.~\eqref{eq:lambda 2 dist} to estimate $\Lambda$ from these two distances, the augmented DEM obtains $\Lambda = 2.02 \pm 0.06$, a $9\%$ improvement over the vanilla RL-optimized DEM's $\Lambda = 1.85 \pm 0.05$.

We also observe that the advantage of a correlation-aware DEM becomes more pronounced at larger code distances: as standard noise mechanisms are suppressed more effectively by the code, the relative impact of unmodeled correlated noise grows, which is reflected in the larger $\Lambda$ obtained with the augmented DEM. The code distances considered in this experimental evaluation, 5 and 7, are smaller than those featured in the simulated results (7 to 15). The larger distances explored in simulation complement the experimental results, offering a glimpse into the regime where quantum error correction is most valuable: where codes are large enough to achieve meaningful logical-error suppression, and where the relative impact of unmodeled noise becomes most pronounced.

\subsection{Comparison with prior work}\label{subsec:prior work}

We now place our work in the context of existing methods. In the introduction, we highlighted the extensive body of work dedicated to optimizing the probabilities of existing error mechanisms in the DEM using experimental data. To our knowledge, however, only two prior works have taken the next step of attempting to learn entirely new error mechanisms from experimental data \cite{arms2025estimating,ye2026reconstruction}.

Both works share a common starting point: rather than building on an existing DEM, they assume no prior knowledge of the device's error mechanisms whatsoever. Arms et al. \cite{arms2025estimating} introduce a cluster-growing algorithm in which a hypergraph is constructed from all pairwise correlations above a cutoff. The algorithm then searches for fully connected components (cliques) in a bottom-up fashion, first identifying all edges (order-2 cliques), then checking whether each edge is contained within a triangle (order-3 clique), and so on up to higher orders. The process terminates either when no higher-order clique is found or when a maximum search order is reached. The resulting cliques form the hyperedges of the decoding hypergraph.  Notably, this permits constructing a new hyperedge out of any subset of detectors, a more general approach than ours, which only combines existing mechanisms.

One important limitation the authors themselves acknowledge is that their method has no natural way to associate logical effects to the newly discovered hyperedges. This prevents the output from being used directly as a prior for decoders, and they identify this as a direction for future work. Ye and Zhang \cite{ye2026reconstruction} use their learned DEM for decoding. However, the mechanism by which logical observable flips are assigned to hyperedges is not detailed, so it remains unclear how this step was resolved in their methodology. Moreover, since their DEM is not decomposed, they are restricted to hypergraph decoders and cannot leverage matching-based decoders, even though they study the surface code.

Our approach differs from both of these works in a key design choice: rather than starting from scratch, we build upon an existing theoretical DEM, such as the SI1000 vanilla DEM, using it as a foundation to focus learning on new, more complex noise mechanisms like correlated noise, instead of relearning basic mechanisms. Perhaps most significantly, grounding new mechanisms in correlations between existing mechanisms provides a natural way to assign logical observables and decompositions to each newly discovered error mechanism, and can therefore naturally be used for decoding. For codes which allow matching-based decoding, this is a substantial advantage of our method: it enables the augmented DEM to be used directly with correlated matching decoders, which require a decomposed DEM, something neither prior work achieves. Beyond this, restricting the candidate set to correlations between known mechanisms substantially reduces the search space, leading to faster convergence and lower sensitivity to statistical noise. Finally, to the best of our knowledge, this is the first time that a method for learning new error mechanisms from experimental data has been shown to yield a measurable improvement in decoding performance on real experimental data.

\section{Summary and outlook}\label{sec:summary}

In this work, we introduced a new method for augmenting an existing DEM by incorporating entirely new error mechanisms that are absent from a theoretically derived model. Our focus was on correlated noise: we developed a general approach capable of identifying and learning any type of correlation between existing error mechanisms directly from experimental data. A key strength of the method is that it assigns a logical effect to each newly discovered error mechanism, the essential step that makes the augmented DEM decodable. It additionally provides a decomposition for each hyperedge, preserving compatibility with graph-based decoders for codes relevant for these decoder. We evaluated the method on the surface code with correlated noise in both simulated and experimental settings. In simulation, for correlation distance 4 the augmented DEM increased $\Lambda$ by $20\%$ relative to the vanilla DEM, nearly matching the oracle DEM's advantage and reducing the logical error rate at distance 15 by more than a factor of 4. For correlation distance 2, the augmented DEM still achieved a $15\%$ increase in $\Lambda$ and more than a threefold reduction in logical error rate at distance 15 relative to the vanilla DEM, though here it did not fully close the gap with the oracle DEM. On experimental data, where only 50,000 samples per circuit were available, we used the recurrence edge method to treat data from different rounds as a substitute for additional samples, and demonstrated a measurable improvement over a strong RL-optimized baseline DEM: a $14\%$ reduction in logical error rate at distance 7, and a $9\%$ increase in $\Lambda$ across distances 5-7. 

Beyond its use in improving decoder priors, we note that the augmentation method can also serve as a diagnostic tool: each discovered hyperedge is tied to an explicit set of detectors and, transitively, physical qubits and circuit locations, together with its estimated strength. This could help localize candidate sources of crosstalk directly from routine QEC data, complementing its primary role in decoding. We leave a fuller exploration of this direction, including attributing hyperedges to specific physical causes, to future work.

Several directions could improve the efficiency of the method going forward. Recall that our augmentation approach considers correlations between any pair of existing mechanisms as candidates, regardless of their physical distance. If greater efficiency is needed, one straightforward optimization is to restrict the candidate set based on physical proximity. Since correlations are expected to occur predominantly between nearby qubits, limiting candidates by the physical distance between error mechanisms could substantially reduce runtime and memory usage. We deliberately avoided this constraint to preserve generality, but it offers a natural optimization for practical deployments.  A complementary and non-mutually exclusive source of efficiency gain is the recurrence edge method (Subsection~\ref{subsec:recurrence}), which was demonstrated in this work for the experimental data. Under the assumption that drift (i.e. changes in the noise over the different rounds) is negligible, pooling recurring instances of a mechanism across rounds reduces the effective number of parameters to be learned and increases the effective sample size, independently of the physical-proximity restriction discussed above. Combining both restrictions, exploiting physical locality together with recurrence across rounds, could further accelerate convergence of the augmentation method beyond what either provides alone.

Finally, as noted in Subsection~\ref{subsec:decoding experimental} and Appendix~\ref{app:residual discrepancy}, applying RL optimization to the weights of all mechanisms in the augmented DEM, rather than relying on the $p_{ij}$ method alone, is a promising avenue for further improving decoding accuracy and closing residual gaps with the oracle DEM.

\bibliographystyle{unsrt}
\bibliography{references}

\appendix

\section{Conditional reweighting in correlation aware decoders}\label{app:adaptivity map}

In this section we describe how decomposed hyperedges are translated into conditional probabilities between their constituent edges, and how these probabilities are used to update weights in the correlated decoding procedure. In this work, we only consider DEMs that decompose into decoding graphs. However, the techniques described here are applicable generally. We recall that a DEM contains a set of independent error mechanisms, their effects on detectors and logical observables, and their associated probability. We use the notation $\mathrm{D}_H = \{(h_i,p_i)\}$ to express the elements in the DEM, where $h_i$ corresponds to an error mechanism (possibly a hyperedge), and $p_i$ is its associated probability. If the DEM admits a decomposition into graphs, then there exists a set of edge and singleton mechanisms $\{e_j\}$ within $\mathrm{D}_H$ such that all elements of $\mathrm{D}_H$ can be expressed using products of $e_j$ \footnote{In some cases, a hyperedge cannot be fully decomposed into existing error mechanisms, leaving a remnant mechanism. If this remnant mechanism is a singleton or an edge, then it is included in the decomposed DEM as a new mechanism. If the remnant is a higher-order hyperedge the decomposition process fails.}. 

We recall that correlated decoding works by performing multiple (usually two) decoder passes. In the first pass, the decoder returns a suggested error $E$. The decoder then uses this correction to conditionally update probabilities on the decoding graph before running a second decoder pass. For each error mechanism identified in the first decoder pass, $e_\mu \in E$, probabilities are updated according to
\begin{equation}
    \Pr(e_\nu |e_\mu) = \frac{\Pr(e_\nu \cap e_\mu)}{\Pr(e_\mu)},
\end{equation}
where $\Pr(e_\mu)$ is obtained by marginalizing over all mechanisms containing $e_\mu$ in the original DEM, $\mathrm{D}_H$, and $\Pr(e_\nu \cap e_\mu)$ is obtained by marginalizing over all mechanisms containing both $e_\nu$ and $e_\mu$ in their decompositions. For independent events, these marginals are equivalent to the probability of an odd number of mechanisms being triggered. That is
\begin{equation}
    \Pr(e_\mu) = \frac{1 - \prod_{e_\mu \in h_i}(1-2p_i)}{2},
\end{equation}
for the single edge, and
\begin{equation}
    \Pr(e_\nu \cap e_\mu) = \frac{1 - \prod_{(e_\nu \cap e_\mu) \in h_i}(1-2p_i)}{2},
\end{equation}
for the joint probability. The updating procedure returns a set of updated probabilities ${e_\nu}$ conditioned upon the first pass $\{e_\mu\}$. If multiple $e_\mu$ from the first pass contribute to the same edge $e_\nu$, the maximum probability is retained. Furthermore, the probability for $e_\nu$ is only updated if the conditional probabilities are larger than the marginal for the first pass, $\Pr(e_\nu)$. The set of conditional probabilities may be computed in advance and used as a conditional reweighting map. Therefore, at runtime, the decoder only needs to update edge weights according to the map and the error string $E$ obtained in the first pass. The second pass then decodes using the updated weights, returning a final correction.

\section{SI1000 noise model}\label{app:si1000}

The SI1000 noise model \cite{gidney2021fault,gidney2022benchmarking,gidney2024inplace} was designed to capture the dominant noise characteristics of superconducting quantum devices. Its appeal lies in its balance between physical realism and analytical tractability. It accounts for a broad range of error mechanisms while remaining fully parameterized by a single global noise strength $p$, which facilitates systematic comparisons across different physical error rates. For these reasons, we use SI1000 as our base noise model throughout the simulations in this work. Parameters for the SI1000 model are shown in Table \ref{tab:SI1000_noise}. Note that noise in this model is restricted to single-qubit channels and two-qubit channels between qubit pairs coupled by a two-qubit gate, with no additional multi-qubit error mechanisms.

\begin{table}[]
    \centering
    \begin{tabular}{c|c|c}
         Operation & Noise channel & Strength  \\
         \hline
         CZ & Two-qubit depolarizing channel & $p$ \\
         One-qubit Clifford gate & One-qubit depolarizing channel & $p/10$ \\
         Reset & One-qubit X channel & $2p$ \\
         Measurement & Measurement error & $5p$ \\
         Idle & One-qubit depolarizing channel & $p/10$ \\
         Resonator Idle & One-qubit depolarizing channel & $2p$
    \end{tabular}
    \caption{Parameters for SI1000 noise \cite{gidney2022benchmarking} used as the base noise model in this work. CZ noise is applied to the two-qubits undergoing the gate. Resonator Idle events correspond to qubits which are idle during a measurement or reset operation.}
    \label{tab:SI1000_noise}
\end{table}

\section{Effect of augmentation cutoffs on simulated decoding performance}\label{app:Simulated threshold}

\begin{figure}
\centering
\includegraphics[scale=0.25]{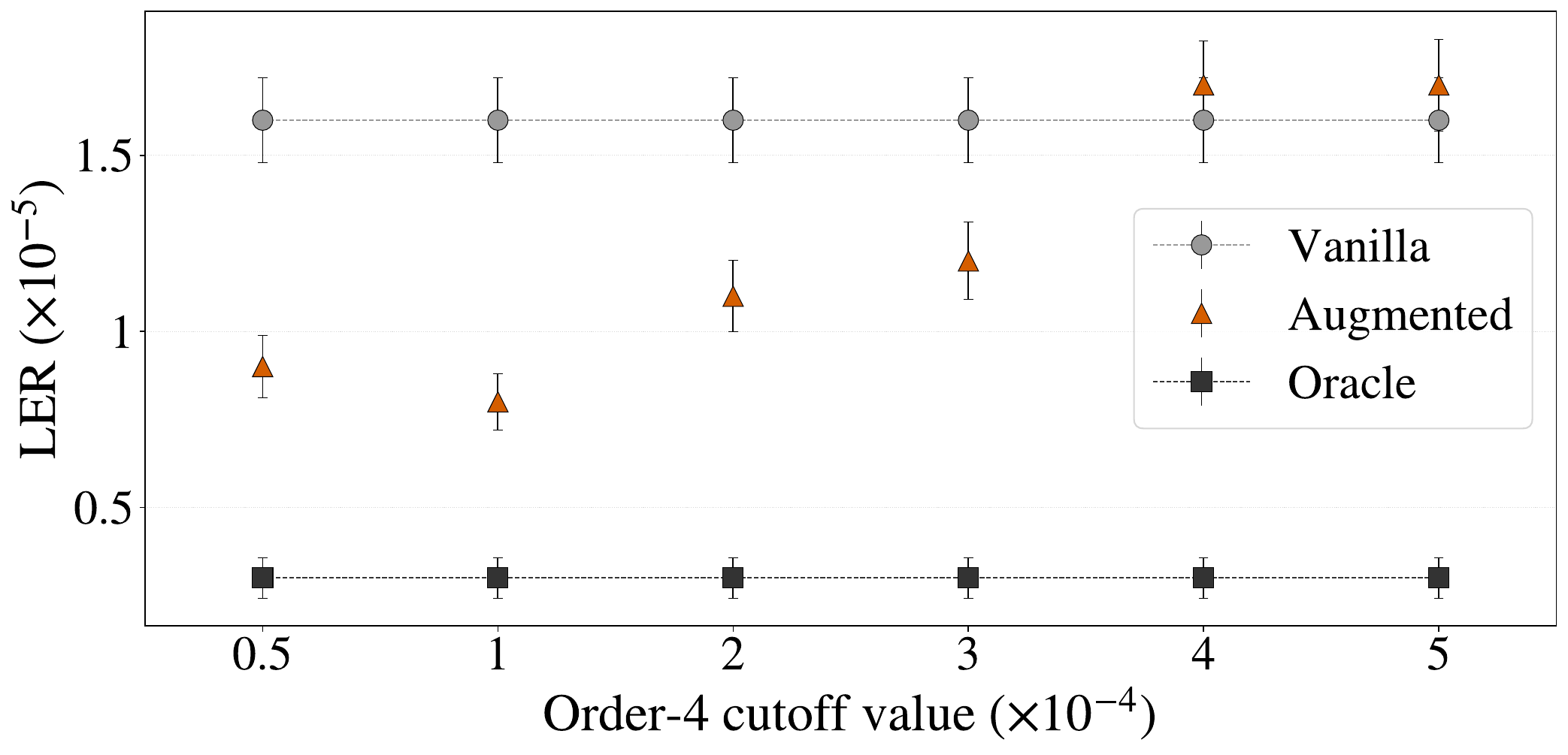}\caption{Effect of augmentation cutoffs parameters on decoding performance for a distance-15 surface code with correlation distance 2, using 50,000 samples for augmentation and a million samples for decoding.  The x-axis of the plot shows the order-4 hyperedge cutoff value $c_4$. The order-3 cutoff, $c_3$, is set  $5\times 10^{-5}$ larger than $c_4$, and the order-2 (edge) cutoff, $c_2$ is set a further $5\times 10^{-5}$ larger than $c_3$. We present the logical error rate (LER) using correlated matching with the oracle (black squares), augmented (orange triangles) and vanilla (gray circles) DEMs. We see that a cutoff as low as $10^{-4}$ for order-4 candidates is more beneficial than larger cutoffs despite the number of spurious mechanisms admitted into the augmentation with this cutoff (see Fig. \ref{fig:recovery plot}).
}\label{fig:threshold sweep}
\end{figure}

To select the cutoff parameters used in Section \ref{sec:results}, we sweep over a range of cutoff values and evaluate the decoding performance on simulated data. Recall that the cutoffs $c_4, c_3$ and $c_2$ control the minimum correlation required for a candidate mechanism to pass the explicit validation, for order-4, order-3, and order-2 candidates respectively. Lower cutoffs admit more candidates, potentially recovering more true mechanisms, but at the cost of introducing more spurious ones and increasing runtime.

Figure \ref{fig:threshold sweep} presents decoding performance across a range of cutoff values, for a distance-15 surface code with correlation distance 2, noise parameters $p=10^{-3}$ and $p_{\mathrm{corr}}=3\times 10^{-4}$, using 50,000 augmentation samples and a million decoding samples. We find that a cutoff as low as $10^{-4}$ for order-4 candidates yields improved decoding accuracy, although, as shown in Figure \ref{fig:recovery plot}, spurious mechanisms substantially outnumber correctly recovered ones at these sample counts. Nevertheless, the inclusion of additional candidates appears to be beneficial on net. Thus we selected $c_4=10^{-4},c_3=1.5\times10^{-4}$ and $c_2=2\times 10^{-4}$ as the cutoffs in our simulations. We also observed similar behavior when applying the recurrence edge method to the experimental data, though there even lower cutoff values proved more beneficial, leading us to select a slightly different set of cutoffs in that case.

\section{LER scaling with augmentation sample number}
\label{app:ler_samples}

\begin{figure}
\centering
\includegraphics[scale=0.24]{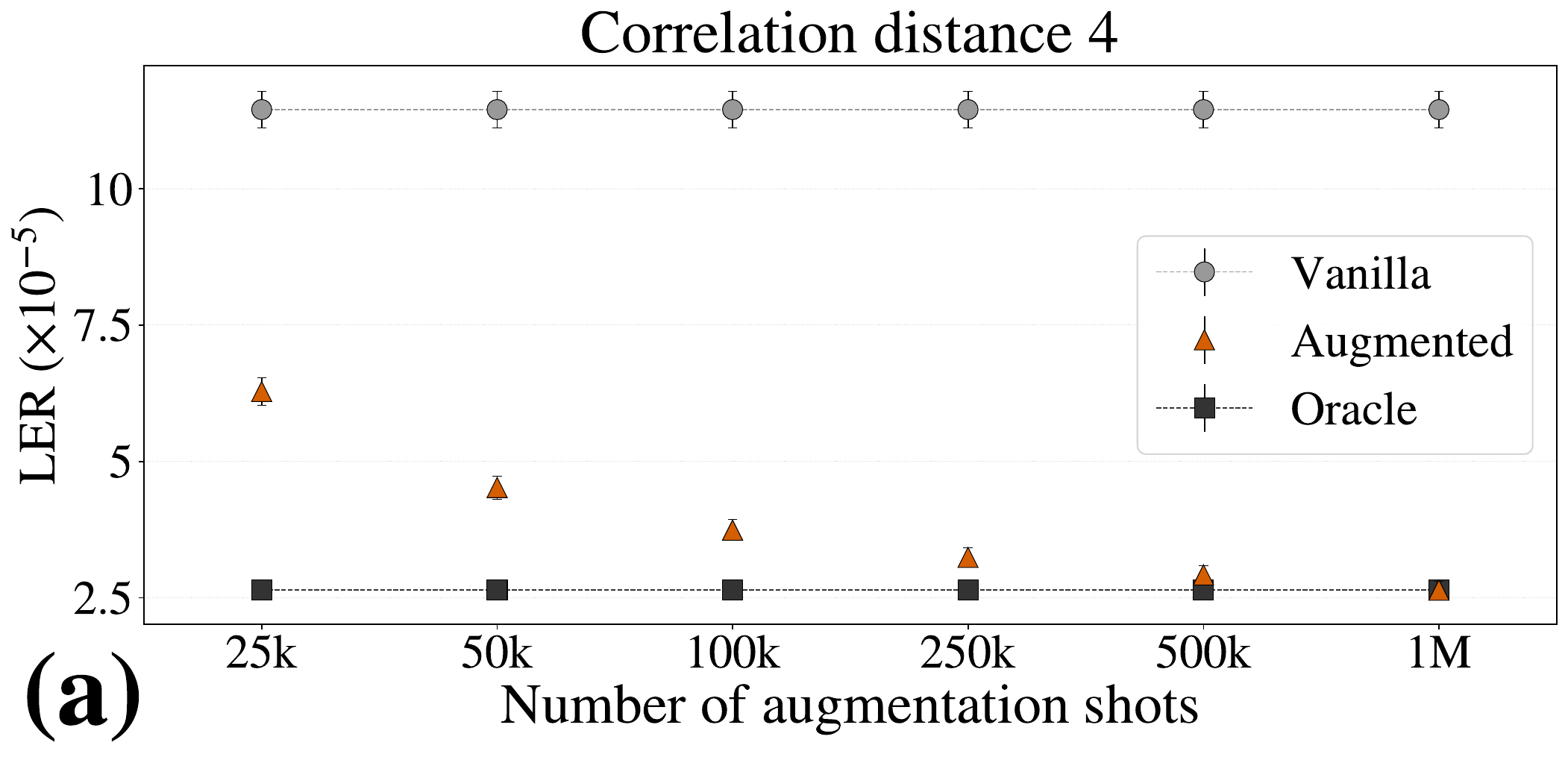}
\hspace{0.6cm}
\includegraphics[scale=0.24]{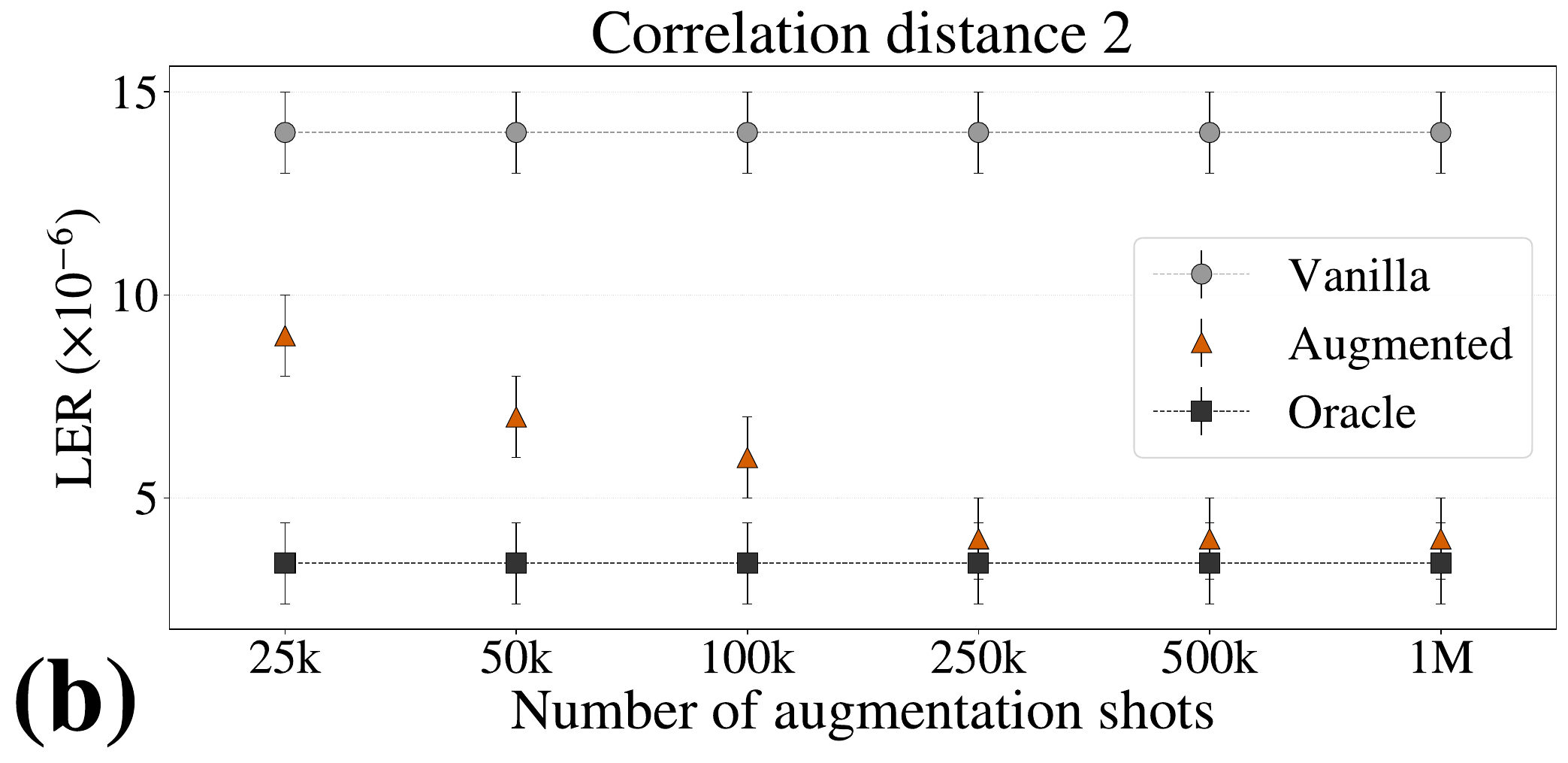}\caption{Decoding performance comparison between the oracle DEM, augmented DEM, and vanilla SI1000 DEM for a surface code of distance 15. The x-axis denotes the number of samples used for augmenting the DEM, while a million samples are used for decoding in all instances. The y-axis denotes the LER obtained when applying correlated decoding with each of the three DEMs, vanilla (gray circles), augmented (orange triangles), oracle (black squares). (a) Correlation distance 4: the augmented DEM fully closes the gap between the vanilla and oracle DEMs at a million augmentation samples. (b) Correlation distance 2: the gap is not fully closed, but the augmented DEM achieves a significant advantage over the vanilla DEM.}\label{fig:decoding plot}
\end{figure}

To complement the main text, Figure~\ref{fig:decoding plot} presents the logical error rate (LER) as a function of the number of augmentation samples. This analysis shows that, for correlation distance 4, the augmentation method nearly fully recovers the oracle DEM's advantage, while for correlation distance 2 the gap is significantly reduced but not fully closed (Appendix \ref{app:residual discrepancy}). Based on this, we select one million augmentation samples for the main-text simulated results in Section \ref{sec:results}, as this is sufficient to recover most of the oracle DEM's advantage. We note that this sample budget could be further reduced by applying the recurrence edge method (Subsection \ref{subsec:recurrence}), which trades additional rounds for additional samples.

\section{Sources and diagnosis of residual discrepancies}\label{app:residual discrepancy}

\subsection{Structural sources of discrepancy}\label{app:discrepancy sources}

Even in the limit of infinitely many samples, where the augmentation method has identified every missing mechanism, certain structural differences between the augmented DEM and the oracle DEM are expected to persist. Here we enumerate the main sources of these discrepancies:
\begin{itemize}

\item Hyperedge decomposition. When a new high-weight error mechanism is identified and we are working with a decomposed DEM, the augmentation method has no way to determine which specific decomposition the oracle DEM uses, since all valid decompositions of a hyperedge are indistinguishable from correlation data alone. The method therefore includes all valid decompositions, whereas the oracle DEM may contain only one. This issue is specific to decomposed DEMs, and the multiple decomposition options arise only for correlation distance 2 in the surface code geometry we consider.

\item Singletons. Certain correlated noise mechanisms give rise to new order-1 error events (singletons), as illustrated in Figure \ref{fig:crosstalk types}(d). We made a deliberate design choice not to search for new singletons in the augmentation method, as they are particularly prone to statistical noise and to contamination from incompletely accounted-for higher-weight mechanisms. As a consequence, these singletons are never discovered, introducing a systematic gap between the augmented and oracle DEMs. Like the decomposition issue, new singletons appear only for correlation distance 2 noise in our simulations.

\item Long-range edges vs. decomposed edges. When there is a correlated noise mechanism between two otherwise-independent detectors, the most natural representation in a decomposed DEM is a decomposed edge, rather than a long-range  undecomposed edge. The augmentation method follows this convention. However, the oracle DEM, generated by Stim directly from the circuit with correlated noise, represents these mechanisms as ordinary edges instead. This discrepancy becomes more significant for larger correlation distances, where very long-range edges may harmfully create very long edges.

\item Unreachable candidates. Depending on the circuit geometry, a small number of new edges in the oracle DEM, introduced by correlated noise mechanisms, may not be obtained by taking products of existing mechanisms in the vanilla DEM. In our observations, we find that such events are rare and localized to the corners of the surface code patch. As the code distance grows, these corner events become negligible compared to the total number of new mechanisms introduced in the bulk. 

\item Vanilla-absorbed correlated mechanisms. There exist certain correlated noise mechanisms whose detector signatures coincide with hyperedges that are already present in the vanilla DEM. This occurs predominantly at correlation distance 2, where the physical proximity of the affected qubits means that the correlation-induced error pattern is indistinguishable, at the detector level, from a standard (non-correlated) error mechanism. Because the augmentation method only adds mechanisms whose detector signatures are absent from the vanilla DEM, these correlated contributions are silently absorbed into existing vanilla entries: the combined probability of the vanilla and correlated mechanisms should be reflected in the DEM, but only the vanilla probability appears. This discrepancy is expected to be resolved naturally once a probability-learning method is applied to the full augmented DEM, as such a method will fit the observed detection-event statistics, which do reflect the correlated contribution, and update the probabilities of existing mechanisms accordingly.

\end{itemize}
Of the sources listed above, only three are specific to correlation distance 2: hyperedge decomposition ambiguity, vanilla-absorbed correlated mechanisms, and missing singletons. In Appendix \ref{subsec:discrepancy plots}, we show that the gap between the oracle DEM LER and the augmented DEM LER is majorly attributable to the first two, while the absence of singletons is not the limiting factor. As discussed, the vanilla-absorbed mechanism discrepancy is readily resolved by a post-processing step that re-learns all probabilities from experimental data after augmentation (e.g. via RL optimization). The hyperedge decomposition ambiguity, on the other hand, is an inherent structural difference that cannot be overcome from experimental data alone. We experimented with alternative methods, such as choosing a single random decomposition instead of including all valid decompositions, and found that this degraded overall performance. It is worth emphasizing that even for correlation distance 2, the augmented DEM still provides a substantial and meaningful improvement over the vanilla SI1000 DEM, as shown in Section \ref{sec:results}.
A possible path to overcoming the impact of this decomposition ambiguity, which is left for future work, is to apply an RL optimization step on the augmented DEM generated by our method, which will treat the probability of each such hyperedge with a valid decomposition as a separate parameter to fine-tune and optimize, hopefully closing the observed discrepancy in the LER.

\subsection{Disentangling Decomposition Ambiguity from Missing Singletons.} \label{subsec:discrepancy plots}

\begin{figure}
\centering

\raisebox{-0.5\height}{\includegraphics[scale=0.34]{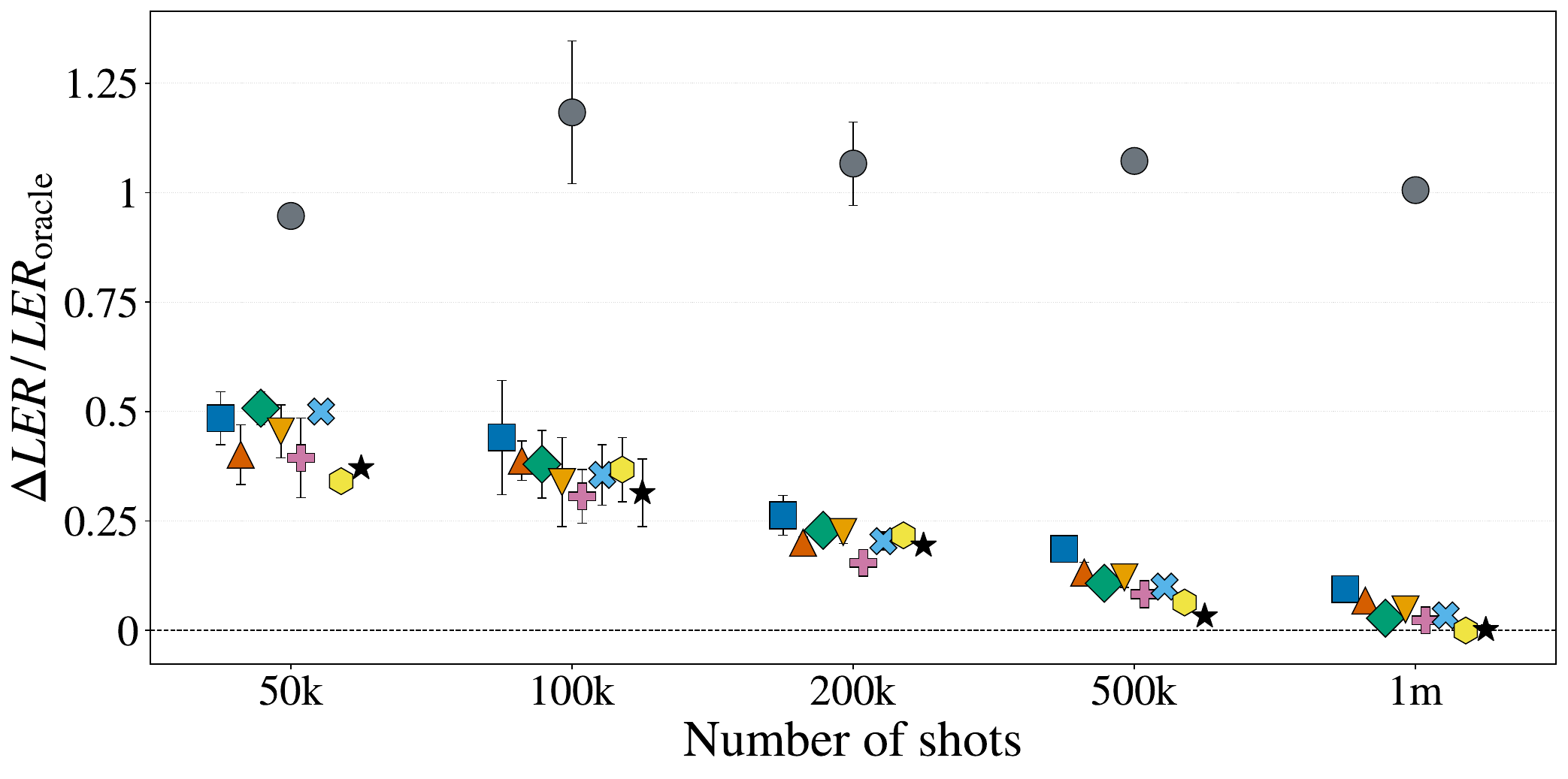}}
\raisebox{-0.4\height}{\includegraphics[scale=0.22]{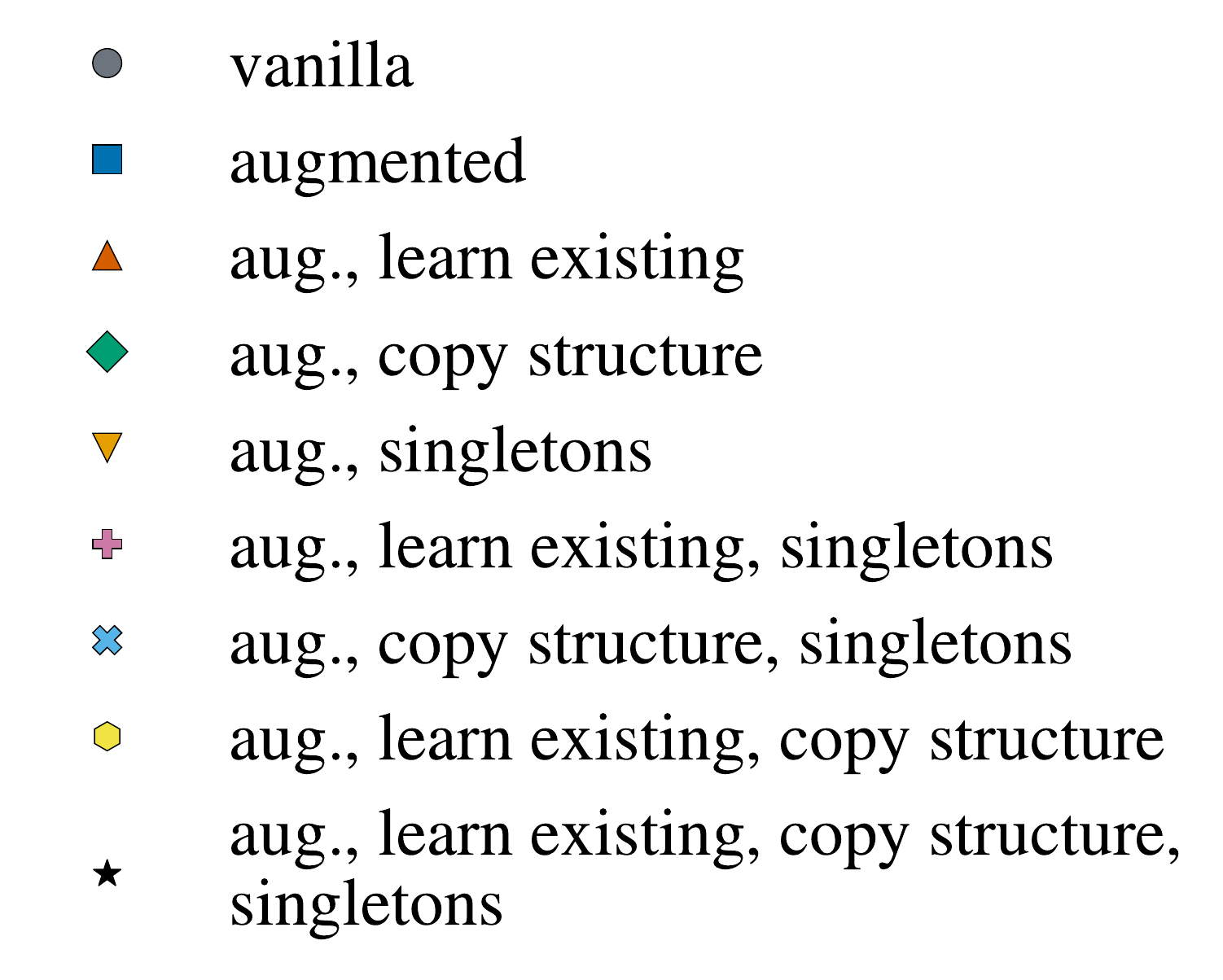}}
\caption{
Analysis of the source of discrepancy between the augmented DEM and the oracle DEM for correlation distance 2, for code distance 7. The x-axis denotes the total number of samples, split equally between augmentation and decoding. Each point shows the relative difference in logical error rate (LER) relative to the oracle DEM. Zero indicates full recovery of the oracle DEM's decoding advantage, while a value of, e.g., $0.5$ means that method's LER is $50\%$ higher than the oracle's. We study three factors that may reduce the remaining gap: (i) learning updated probabilities for mechanisms already present in the vanilla DEM from the data (``learn existing''), (ii) copying the hyperedge decomposition structure of recovered mechanisms directly from the oracle DEM into the augmented DEM (``copy structure''), and (iii) copying all singleton error mechanisms absent from the vanilla DEM directly from the oracle DEM into the augmented DEM (``singleton''). Each series in the legend corresponds to a distinct combination of one or more of these factors applied to the augmented DEM. For reference, results for the vanilla DEM (gray circles) and the unmodified augmented DEM (blue squares) are also shown.
Small horizontal shifts are applied to prevent marker overlap and are purely a visual aid, carrying no additional meaning. This figure demonstrates that, at least for this example, these three factors together fully account for the gap, and notably, adding singletons is not even required to close it.
}\label{fig:modified augmented dem}
\end{figure}

To investigate the sources of the performance gap observed in Figure \ref{fig:decoding plot}, we repeat the decoding experiment with the augmented DEM under a series of controlled modifications, each targeting one of the suspected causes. Figure \ref{fig:modified augmented dem} shows the discrepancy in LER from the oracle DEM for each controlled modification when simulating a distance-7 experiment with correlation distance 2 noise. Also included for reference are results for the unmodified augmented DEM and vanilla DEM.

The first modification is to update the probabilities of error mechanisms that overlap with ones already present in the vanilla DEM, in addition to adding the new candidate mechanisms discovered by the augmentation method. Unlike the modifications that follow, this one can genuinely be incorporated into the augmentation method whenever it improves accuracy, and it was indeed adopted in Subsection \ref{subsec: results simulated} for correlation distance 2. We refer to this as ``learn existing'' in the legend. Applied on its own, it is marked by a red triangle in Figure \ref{fig:modified augmented dem}, where it visibly improves decoding accuracy without fully closing the gap. This modification addresses the ``Vanilla-absorbed correlated mechanisms'' issue discussed in Appendix \ref{app:discrepancy sources}.

The second modification we study is copying the hyperedge decomposition structure from the oracle DEM for all new error mechanisms found by the augmentation method, addressing the ``hyperedge decomposition'' and ``long-range edges vs. decomposed edges'' issues from Appendix \ref{app:discrepancy sources}. This cannot actually be performed by the augmentation method itself and is used here purely as a diagnostic tool to isolate the remaining sources of discrepancy. We call this ``copy structure'' in the legend. On its own, it is marked by a green diamond in Figure \ref{fig:modified augmented dem}.

The third modification is adding all singleton mechanisms missing from the augmented DEM directly from the oracle DEM, addressing the ``singletons'' issue from Appendix \ref{app:discrepancy sources}. In principle, these missing singletons could be learned by the augmentation method. However, since the probabilities of single-detector mechanisms are highly susceptible to statistical fluctuations, many spurious high-probability candidates tend to appear, and including them was not found to improve performance. We therefore do not attempt to learn the singletons from data, but instead copy them directly from the oracle DEM, again purely as a diagnostic tool, to quantify how much accuracy is lost by their absence from the augmented DEM. We call this ``singletons'' in the legend. On its own, it is marked by an orange inverted triangle in Figure \ref{fig:modified augmented dem}.

While each of these modifications individually improves the LER, none fully closes the gap on its own, motivating us to examine combinations. Combining ``learn existing'' with ``singletons'' (pink plus sign) and combining ``copy structure'' with ``singletons'' (light blue x) each yield further improvement, but still fall short of closing the gap. The first combination to fully close the gap is ``learn existing'' together with ``copy structure'' (yellow hexagon). Adding ``singletons'' on top of these two (black star) achieves the same result. This indicates that, at least for this example, these three factors together fully account for the gap, and notably, adding singletons is not even required to close it.

We also note that Stim routinely generates DEMs containing duplicate error instructions, i.e., multiple instructions sharing the same detector sets and decompositions but potentially differing in probability. To simplify these modifications, specifically copying hyperedge decompositions from the oracle DEM, we eliminate such duplicates as a preprocessing step applied before any augmentation or decoding. We first flatten all circuits prior to constructing the DEMs, as repeat blocks were found to be the primary source of duplicates. Any remaining duplicates are then merged into a single instruction with combined probability $1-(1-p_1)(1-p_2)$ and recursively so for higher multiplicities.
\end{document}